%% file: Main.tex
\documentclass[nofootinbib]{article}

\usepackage{amssymb}

\usepackage{amsmath}

\usepackage{bm}

\usepackage{jcappub}

\extrafloats{72}

\include{mydefs}

\title{Apples with Apples comparison of 3+1 conformal numerical relativity schemes}

\author{David Daverio$^1$,} 
\emailAdd{dd415@damtp.cam.ac.uk}
\author{Yves Dirian$^{2,3}$,}
\emailAdd{yves.dirian@unige.ch}
\author{Ermis Mitsou$^3$}
\emailAdd{ermitsou@physik.uzh.ch}

\affiliation{$^1$\,Centre for Theoretical Cosmology, Department of Applied Mathematics and Theoretical Physics, Wilberforce Road, Cambridge CB3 0WA, United  Kingdom}
\affiliation{$^2$\,Department of Theoretical Physics and Center for Astroparticle Physics, University of Geneva, 24 quai Ansermet, CH--1211 Gen\`eve 4, Switzerland}
\affiliation{$^3$\,Center for Theoretical Astrophysics and Cosmology, Institute for Computational Science, University of Z\"urich, CH--8057 Z\"urich, Switzerland}

\abstract{This paper contains a comprehensive comparison catalog of ``Apples with Apples'' tests for the BSSNOK, CCZ4 and Z4c numerical relativity schemes, with and without constraint damping terms for the latter two. We use basic numerical methods and reach the same level of accuracy as existing results in the literature. We find that the best behaving scheme is generically CCZ4 with constraint damping terms.}

\begin{document}

\maketitle

\flushbottom

\section{Introduction}

The most widely employed approaches for solving numerically the Einstein equations, especially in the context of binary black hole simulations, are currently the Generalized Harmonic scheme \cite{HF, Garfinkle, SW, Pretorius, LSKOR} and the BSSNOK scheme \cite{NOK, SN, BS}. With the advent of the ``Z4" formulation of the Einstein equations \cite{BLPZ, GMGCH}, whose primary aim is to improve the control over constraint violation, two Z4-based extensions of BSSNOK have been considered in the literature: the ``Z4c" scheme \cite{BH,WBH} and the ``CCZ4" scheme \cite{ABCBRP}. The aim of this paper is to provide a detailed comparison of the BSSN, Z4c and CCZ4 schemes based on the minimal testbed that was laid out in \cite{A2A1, A2A2} and is commonly referred to as the ``Apples with Apples" tests (AwA). 

Part of our motivation stems from the fact that the literature currently lacks a comprehensive set of AwA results, even for the most prominent schemes in the field such as BSSNOK. This information could serve as a reference for comparing one's code with and would therefore be especially useful to new researchers in the field. In the case where one is interested in comparing schemes, it is also important to have the tests run using the same code and numerical methods, so as to obtain the cleanest possible comparison of the schemes' true merits. Occurrences of such comprehensive AwA treatments are for instance found in \cite{A2A1, A2A2} and in \cite{CH} for the Z4c scheme, but with second-order time integration, contrary to the present standard that is fourth-order (see also the recent \cite{DGKRZ} where a more sophisticated discretization technique is used). Here we provide a relatively exhaustive catalog of AwA results, mentioning all the relevant information and using several standard gauge choices when this freedom is present. We use ``vanilla" numerical methods and reach the same level of accuracy as the few results we found in the literature in the case of similar discretization techniques \cite{CFFKLT, MGS}. 

We also take advantage of this flurry of data to propose an alternative analysis of wave-based tests, such as the linear and gauge wave tests for instance. Instead of focusing on the wave profile and its error in real space, as suggested in \cite{A2A1, A2A2}, we consider the discrete Fourier transform at each crossing time, because it allows us to disentangle the errors on the wave's phase, amplitude and offset and also provides a direct measurement of higher mode contamination. 

As for the results of the comparison, we find that the ``damped" CCZ4 scheme behaves generically better than the rest. Given the proximity between the CCZ4 and Z4c formulations, it is interesting to pinpoint the distinguishing property underlying the difference in their performance. For this reason, we have included a fourth scheme in our comparison, which we dubbed ``Z4cc", that lies ``in between" CCZ4 and Z4c. More precisely, Z4cc uses the same redefinition of the extrinsic curvature trace $K$ as Z4c, when deriving it from the Z4 equations, but unlike Z4c, no pure-constraint terms are discarded, thus making it ``covariant" like CCZ4. It turns out that the Z4cc results are practically indistinguishable from the CCZ4 ones, and thus not displayed here, which means that the crucial property is covariance and not whether one redefines $K$.

The paper is organized as follows. In section \ref{sec:schemesgauges} we present the schemes that will be compared, i.e. CCZ4, Z4cc, Z4c, both in their damped and undamped versions, and two versions of BSSNOK (covariant and non-covariant), and we also discuss the gauge choices we will consider when possible. In section \ref{sec:specs} we provide general specifications, including the numerical methods that are used. Each of the subsequent sections is then devoted to a particular AwA test, starting with its description and followed by comments about the results. All figures are collected at the end of the paper.

This is a abridged version with a selected sublist of plots. The full version can be found: \href{https://www.dropbox.com/s/wuu5k8byuz8ectw/AwA_full.pdf?dl=0}{{\tt full version}}.

\section{Schemes and gauges} \label{sec:schemesgauges}

For the purpose of this paper it is enough to begin with the Z4 formulation of vacuum GR \cite{BLPZ} with constraint damping term \cite{GMGCH}
\beq \label{eq:Z4}
R_{\mu\nu} + \na_{\mu} Z_{\nu} + \na_{\nu} Z_{\mu} - \ka \[ n_{\mu} Z_{\nu} + n_{\nu} Z_{\mu} - g_{\mu\nu} n_{\ro} Z^{\ro} \] = 0 \, ,
\eeq
where we have set the secondary constraint-damping parameter of \cite{GMGCH} to zero for simplicity. From the $3+1$ viewpoint, the $Z_{\mu}$ field must be assigned zero initial conditions and this value is then preserved under time-evolution at the analytical level. It therefore constitutes an extra set of constraints, i.e. on top of the Hamiltonian and momentum ones, and can be thought of as a conjugate variable of the latter. 

We first perform a $3+1$ decomposition, i.e. the line-element becomes
\beq
\ed s^2 = -\al^2 \ed t^2 + \ga_{ij} \( \ed x^i + \be^i \ed t \) \( \ed x^j + \be^j \ed t \) \, , 
\eeq
and also define
\beq
\Te := - n_{\mu} Z^{\mu} \, , \hspace{1cm} K_{ij} := - \frac{1}{2\al} \( \pa_t - \Lie_{\be} \) \ga_{ij} \, ,
\eeq
where $\Lie_{\be}$ is the Lie derivative with respect to $\be^i$. In the plots we will display, we will often use the notation ``$g_{ij}$" instead of ``$\ga_{ij}$" for the 3-metric, since it is more common to all numerical relativity schemes. We next perform a conformal/traceless decomposition
\beq
\ga_{ij} = \chi^{-1} \ti{\ga}_{ij} \, , \hspace{1cm} \det \ti{\ga}_{ij} \equiv 1 \, ,
\eeq
\beq
K_{ij} = \chi^{-1} \[ \ti{A}_{ij} + \frac{1}{3}\, \ti{\ga}_{ij} K \] \, , \hspace{1cm} \ti{\ga}^{ij} \ti{A}_{ij} \equiv 0 \, ,
\eeq
and use the convention of displacing the indices of tilded tensors with $\ti{\ga}_{ij}$. Finally, we trade $Z_i$ for 
\beq
\hat{\Ga}^i := 2 \ti{\ga}^{ij} Z_j + \ti{\Ga}^i \, , \hspace{1cm} \ti{\Ga}^i := - \pa_j \ti{\ga}^{ij} \, .
\eeq
The equations \eqref{eq:Z4} now become a set of evolution equations for $\Te$, $\chi$, $K$, $\hat{\Ga}^i$, $\ti{\ga}_{ij}$ and $\ti{A}_{ij}$. Here we consider a more general class of equations depending on two parameters $s$ and $c$ taking values in the set $\{ 0, 1 \}$, with the \eqref{eq:Z4} case corresponding to $(s,c) = (0,1)$. These are 
\bea
\ced_t \Te & = & \al \[ \frac{1}{3}\, K^2 - \( 1 - \frac{4}{3}\, s \) K \Te - \ka \Te - \frac{2}{3}\, s \Te^2 - \frac{1}{2}\, \ti{A}_{ij} \ti{A}^{ij} + \frac{1}{2}\, \ti{\ga}^{ij} \hat{R}_{ij} \] + \ti{Z}^i \( \al \pa_i \chi - \chi \pa_i \al \)   \, , \\
\ced_t \chi & = & \frac{2}{3}\, \chi \[ \al \( K + 2 s \Te \) - \pa_i \be^i \] \, , \\
\ced_t K & = & \al \[ \( 1 - \frac{2}{3}\, s \) K^2 - 2 \( 1 - \frac{5}{3}\, s \) K \Te - \( 3 - 4s \) \ka \Te + \frac{4}{3}\, s \Te^2 + s \ti{A}_{ij} \ti{A}^{ij} + 2 \( 1 - s \) c \ti{Z}^i \pa_i \chi \] \nn \\
 & & +\, \chi \hat{\Ga}_{sc}^i \pa_i \al  + \ti{\ga}^{ij} \[ - \chi \pa_i \pa_j \al + \frac{1}{2}\, \pa_i \chi \pa_j \al + \al \( \( 1 - s \) \hat{R}_{ij} + \frac{1}{2}\, \chi S_{ij} \) \] \, , \\
\ced_t \hat{\Ga}^i & = & 2\al \[ \ti{\Ga}^i_{jk} \ti{A}^{jk} - \frac{3}{2}\, \ti{A}^{ij} \chi^{-1} \pa_j \chi + \ti{\ga}^{ij} \( \( 1 - \frac{4}{3}\, s \) \pa_j \Te - \frac{2}{3}\, \pa_j K \) - c\( \frac{2}{3} \, K + \frac{4}{3}\,s \Te + \ka \) \ti{Z}^i \] \nn \\
 & & - 2c \Te \ti{\ga}^{ij} \pa_j \al - 2 \ti{A}^{ij} \pa_j \al + \ti{\ga}^{jk} \pa_j \pa_k \be^i + \frac{1}{3}\, \ti{\ga}^{ij} \pa_j \pa_k \be^k  - \hat{\Ga}_c^j \pa_j \be^i + \frac{2}{3}\, \hat{\Ga}_c^i \pa_j \be^j \, , \label{eq:DGamma} \\
\ced_t \ti{\ga}_{ij} & = & - 2 \al \ti{A}_{ij} + \ti{\ga}_{ik} \pa_j \be^k + \ti{\ga}_{jk} \pa_i \be^k - \frac{2}{3}\, \ti{\ga}_{ij} \pa_k \be^k \, , \\
\ced_t \ti{A}_{ij} & = & \al \[ - 2 \ti{\ga}^{kl} \ti{A}_{ik} \ti{A}_{jl} + \( K - 2 \( 1 - s \) \Te \) \ti{A}_{ij} \] + \ti{A}_{ik} \pa_j \be^k + \ti{A}_{jk} \pa_i \be^k - \frac{2}{3}\, \ti{A}_{ij} \pa_k \be^k \nn \\
 & & + \[ \chi \( - \pa_i \pa_j \al + \ti{\Ga}^k_{ij} \pa_k \al \) - \pa_{(i} \chi \pa_{j)} \al + 2 c\al \ti{Z}^k \ti{\ga}_{k(i} \pa_{j)} \chi + \al \( \hat{R}_{ij} - \chi S_{ij} \)  \]^{\rm TF}  \, ,
\eea 
where ``TF" denotes the trace-free part and we have defined
\bea
\ced_t & := & \pa_t - \be^i \pa_i \, , \\
\ti{Z}^i & := & \frac{1}{2} \[ \hat{\Ga}^i - \ti{\Ga}^i \] \, , \\
\hat{\Ga}_q^i & := & q \hat{\Ga}^i + \( 1 - q \) \ti{\Ga}^i \, , \\
\ti{\Ga}_{kij} & := & \frac{1}{2} \( \pa_i \ti{\ga}_{jk} + \pa_j \ti{\ga}_{ik} - \pa_k \ti{\ga}_{ij} \) \, , \\
\ti{\Ga}^k_{ij} & := & \ti{\ga}^{kl} \ti{\Ga}_{lij} \, , \\
\hat{R}_{ij} & := & R_{ij}[\ga] + 2 \na^{\ga}_{(i} Z_{j)} - 2 c Z_{(i} \pa_{j)} \log \chi  \nn \\
 & \equiv & \chi \[ - \frac{1}{2}\, \ti{\ga}^{kl} \pa_k \pa_l \ti{\ga}_{ij} + \ti{\ga}_{k(i} \pa_{j)} \hat{\Ga}^k + \ti{\Ga}_{(ij)k} \hat{\Ga}_c^k + \ti{\ga}^{kl} \( \ti{\Ga}^m_{ki} \ti{\Ga}_{mlj} + 2\ti{\Ga}^m_{k(i} \ti{\Ga}_{j)ml} \) \] \label{eq:Rhat}  \\
 & & +\, \frac{1}{2} \[ \pa_i \pa_j \chi - \frac{1}{2}\, \chi^{-1} \pa_i \chi \pa_j \chi + \ti{\ga}_{ij} \ti{\ga}^{kl} \( \pa_k \pa_l \chi - \frac{3}{2}\, \chi^{-1} \pa_k \chi \pa_l \chi \) - \ti{\Ga}^k_{ij} \pa_k \chi - \ti{\ga}_{ij} \hat{\Ga}^k_c \pa_k \chi  \] \, . \nn
\eea
The constraint equations are
\bea
D & := & \det \ti{\ga} - 1 = 0 \, , \label{eq:Dconstraint} \\
D' & := & \ti{\ga}^{ij} \ti{A}_{ij} = 0 \, , \label{eq:Tconstraint} \\  
\Te & = & 0 \, , \\
\ti{Z}^i & := & \frac{1}{2} \[ \hat{\Ga}^i - \ti{\Ga}^i \] = 0 \, , \\
H & := & - \frac{1}{3}\, K^2 + \frac{1}{2}\, \ti{A}_{ij} \ti{A}^{ij} - \frac{1}{2}\, \ti{\ga}^{ij} \hat{R}_{ij} - c \ti{Z}^i \pa_i \chi = 0 \, , \nn \\
\\
M_i & := & - \ti{\ga}^{jk} \[ \pa_j \ti{A}_{ki} - \ti{A}_{li} \ti{\Ga}^l_{kj} - \ti{A}_{kl} \ti{\Ga}^l_{ij} - \frac{3}{2}\, \ti{A}_{ij} \chi^{-1} \pa_k \chi \] + \frac{2}{3}\, \pa_i K = 0 \, ,
\eea
where the last two are the vacuum Hamiltonian (up to a $\na_i Z^i$ term) and momentum constraints. 

Setting $c = 0$ amounts to using the $\Te, \ti{Z}^i = 0$ constraints in several parts of the equations. Following the terminology of \cite{ABCBRP}, these schemes are therefore no longer ``covariant", since non-4-covariant equations have been used. On the other hand, setting $s = 1$ amounts to the redefinition
\beq
K \to K + 2 \Te \, .
\eeq 
We can now distinguish the eight different schemes that will be compared in this paper
\begin{itemize}
\item $s = 0$, $c = 1$, $\ka = 0$ \hspace{1.84cm} $\Rightarrow$ \hspace{0.2cm} ``CCZ4-u" scheme \cite{ABCBRP}
\item $s = 0$, $c = 1$, $\ka = L^{-1}$ \hspace{1.41cm} $\Rightarrow$ \hspace{0.2cm} ``CCZ4-d" scheme \cite{ABCBRP}
\item $s = 1$, $c = 1$, $\ka = 0$ \hspace{1.84cm} $\Rightarrow$ \hspace{0.2cm} ``Z4cc-u" scheme 
\item $s = 1$, $c = 1$, $\ka = L^{-1}$ \hspace{1.41cm} $\Rightarrow$ \hspace{0.2cm} ``Z4cc-d" scheme 
\item $s = 1$, $c = 0$, $\ka = 0$ \hspace{1.84cm} $\Rightarrow$ \hspace{0.2cm} ``Z4c-u" scheme \cite{BH}
\item $s = 1$, $c = 0$, $\ka = L^{-1}$ \hspace{1.41cm} $\Rightarrow$ \hspace{0.2cm} ``Z4c-d" scheme \cite{WBH}
\item $s = 1$, $c = 1$, $\ka = 0$, $\Te \equiv 0$ \hspace{0.74cm} $\Rightarrow$ \hspace{0.2cm} ``BSSNcc" scheme \cite{NOK, SN, BS}
\item $s = 1$, $c = 0$, $\ka = 0$, $\Te \equiv 0$ \hspace{0.74cm} $\Rightarrow$ \hspace{0.2cm} ``BSSNc" scheme \cite{NOK, SN, BS} 
\end{itemize}
where $L$ is a length scale of relevance in the simulation at hand. The $c = 0$ schemes have only one ``c" in their name, standing for ``conformal", whereas the $c = 1$ schemes have two ``c" because they are both conformal and covariant. In particular, the BSSNc and BSSNcc schemes are two usual versions of the well-known BSSNOK scheme. As for the ``u" and ``d" letters, they stand for ``undamped" and ``damped", respectively, i.e. depending on the value of the constraint-damping parameter $\ka$. Note that we have also added a new (to our knowledge) scheme, which is the covariant analogue of the Z4c scheme, hence dubbed ``Z4cc". The aim of including this scheme is to determine whether the differences between the CCZ4 and Z4c schemes are due to the choice of $c$ parameter or to the choice of $s$ parameter. As we will see, Z4cc is qualitatively indistinguishable from CCZ4, and both behave generically better than Z4c, meaning that the important property is covariance, i.e. $c = 1$.

Finally, here are the gauge choices that will be considered in this paper. For the slicing we will use the Bona-Mas\'o class \cite{BMSS}
\beq
\ced_t \al = - \al^2 f(\al) \[ K - 2 \( 1 - s \) \Te \] \, ,
\eeq
and, more precisely, the geodesic slicing $f = 0$ (GEO), harmonic slicing $f = 1$ (HARM) and ``$1+\log$" slicing $f = 2/\al$ (LOG). For the shift we will consider the case $\be^i = 0$ (ZERO), the following hyperbolic ``Gamma-driver" case (DRIVER) \cite{BCCKvM,CLMZ}
\beq
\ced_t \be^i = \frac{3}{4}\, B^i  \, , \hspace{1cm} \ced_t B^i = \ced_t \hat{\Ga}^i - B^i  \, ,
\eeq
and the ``harmonic" case (HARM)
\beq
\ced_t \be^i = \al \[ \ti{\ga}^{ij} \( - \chi \pa_j \al + \frac{1}{2}\, \al \pa_j \chi \) + \al \chi \hat{\Ga}^i \] \, .
\eeq
The latter actually corresponds to the true harmonic gauge, up to a $\ti{Z}^i$ term, only if we also choose harmonic slicing. With three possible slicings and three possible shift choices, we have nine possible gauge combinations. Among these choices, the LOG-DRIVER gauge stands out as the one leading to successful binary black hole simulations \cite{BCCKvM,CLMZ} when the singularity is treated using the ``puncture" method \cite{BB}. One should therefore pay special attention to the performance of the scheme in that gauge for the tests that allow one to choose it, namely, the robust stability and the linear wave tests.

\section{General specifications} \label{sec:specs}

Following \cite{A2A2}, all tests are performed in a box of length $L = 1$ with $N = 50 \ro$ points, for the three resolutions $\ro \in \{ 1, 2, 4 \}$. The lattice spacing is therefore $\De x := L/N = 0.02 \ro^{-1}$ and the constraint-damping parameter is $\ka = 1$ when different than zero. We also let $T$ denote the number of crossing times for which a given simulation is running.

For the spatial derivatives we use a centered five-point stencil, i.e.
\beq
\pa_i f(\vec{x}) \to \frac{-2 f(\vec{x} + 2\vec{\De}_i) + 8f(\vec{x} + \vec{\De}_i) - 8f(\vec{x} - \vec{\De}_i) + 2 f(\vec{x} - 2\vec{\De}_i)}{12 \De x} \, , 
\eeq
where
\beq
(\vec{\De}_i)_j := \de_{ij} \De x \, .
\eeq
There are two exceptions to this. First, the $\pa_i$ appearing inside the convective derivative $\ced_t := \pa_t - \be^i \pa_i$ is replaced with the up/down-wind five-point stencil, depending on the sign of $\be^i$ 
\beq
\be^i \pa_i f(\vec{x}) \to \frac{\be^i}{12 \De x} \times \left\{ \begin{array}{ccc} f(\vec{x} + 3\vec{\De}_i) - 6 f(\vec{x} + 2\vec{\De}_i) + 18 f(\vec{x} + \vec{\De}_i) - 10 f(\vec{x}) - 3 f(\vec{x} - \vec{\De}_i) & {\rm if} & \be^i > 0 \\ - f(\vec{x} - 3 \vec{\De}_i) + 6 f(\vec{x} - 2 \vec{\De}_i) - 18 f(\vec{x} - \vec{\De}_i) + 10 f(\vec{x}) + 3 f(\vec{x} + \vec{\De}_i) & {\rm if} & \be^i < 0  \end{array} \right.  \, .
\eeq
Second, whenever we have double derivatives $\pa_i \pa_j$, the diagonal terms are replaced with the second derivative centered five-point stencil
\beq
\pa_i \pa_i f(\vec{x}) \to \frac{-f(\vec{x} + 2\vec{\De}_i) + 16f(\vec{x} + \vec{\De}_i) - 30f(\vec{x}) + 16f(\vec{x} - \vec{\De}_i) - f(\vec{x} - 2\vec{\De}_i)}{12 \De x^2} \, .
\eeq
As for the time integration, we use the fourth-order Runge-Kutta method with time-step $\De t := C \De x$, so that $C$ is the Courant factor. Moreover, we impose the constraint \eqref{eq:Tconstraint} by hand 
\beq
\ti{A}_{ij} \to \ti{A}_{ij} - \frac{1}{3}\, \ti{\ga}_{ij} \ti{\ga}^{kl} \ti{A}_{kl} \, ,
\eeq
after every Runge-Kutta {\it sub-}step. We also include Kreiss-Oliger dissipation \cite{KO}, choosing the sixth-order one since we use a fourth-order time-integration. Thus, to each evolution equation we append the term
\bea
\dot{f}(\vec{x}) & \to & \dot{f}(\vec{x}) + \frac{\si}{64 \De x} \sum_{i=1}^3 \[ f(\vec{x} + 3\vec{\De}_i) - 6 f(\vec{x} + 2\vec{\De}_i) + 15 f(\vec{x} + \vec{\De}_i) - 20 f(\vec{x}) \right. \nn \\
 & & \hspace{2.5cm} \left. +\, 15 f(\vec{x} - \vec{\De}_i) - 6 f(\vec{x} - 2\vec{\De}_i) + f(\vec{x} - 3\vec{\De}_i)  \] \, .  
\eea
The normalization of the parameter $\si$ is such that the stability bounds are \cite{Alcubierre}
\beq
0 \leq \si \leq 2 C^{-1} \, .
\eeq
Finally, the initial conditions of $\hat{\Ga}^i$ are given by the analytical solution of $- \pa_j \ti{\ga}^{ij}$, {\it not} by taking numerical spatial derivatives of the $\ti{\ga}_{ij}$ field on the grid. Also, given some integrated field $X$, we denote the absolute error between the numerical solution and the analytical one by
\beq
\De X := |X_{\rm num.} - X_{\rm an.}| \, .
\eeq

\section{Robust stability test}

\subsection{Specifications}

The initial conditions are given by
\bea
\al, \chi = 1 + \ep \, , \hspace{1cm} K, \Te = \ep \, , \hspace{1cm} \hat{\Ga}^i, \be^i, B^i = \ep^i \, , \hspace{1cm} \ti{\ga}_{ij} = \de_{ij} + \ep_{ij} \, , \hspace{1cm} \ti{A}_{ij} = \ep_{ij} \, ,
\eea
where the $\sim \ep_{\dots}$ fields on the right-hand side are randomly drawn in the interval 
\beq
\ep_{\dots} \in [-10^{-10}/\ro^2, 10^{-10}/\ro^2] \, ,
\eeq
independently for every component and for every grid point. The parameters are set to 
\beq
T = 10 \, , \hspace{1cm} C = 0.1 \, , \hspace{1cm} \si = 0 \, . 
\eeq
We perform the test for all nine gauge choices and plot the evolution of the $L_{\infty}$ norm of $g_{ij} - \de_{ij}$, $\Te$, $\ti{Z}^i$, $H$ and $M_i$.

\subsection{Comments}

\begin{figure}
\includegraphics[width=\columnwidth]{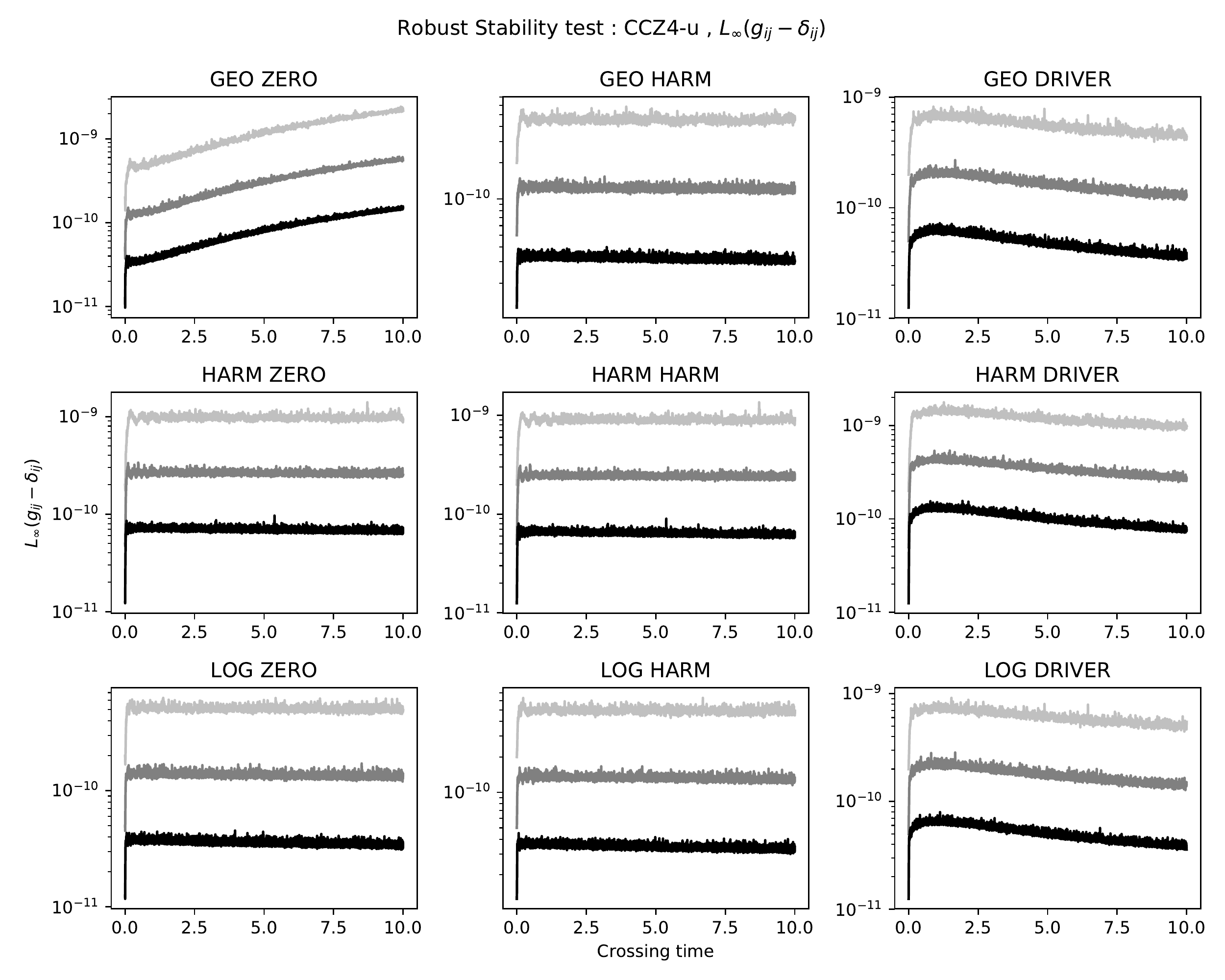} 
\caption{The $L_{\infty}$ norm of $g_{ij} - \de_{ij}$ as a function of the number of crossing times in the robust stability test of the CCZ4-u scheme for all nine gauges choices. Each plot contains the three resolutions $\ro \in \{ 1,2,4 \}$ corresponding to the light-gray, dark-gray and black lines, respectively.}
\label{fig:Rob_ga_CCZ4u}
\end{figure}

\begin{figure}
\includegraphics[width=\columnwidth]{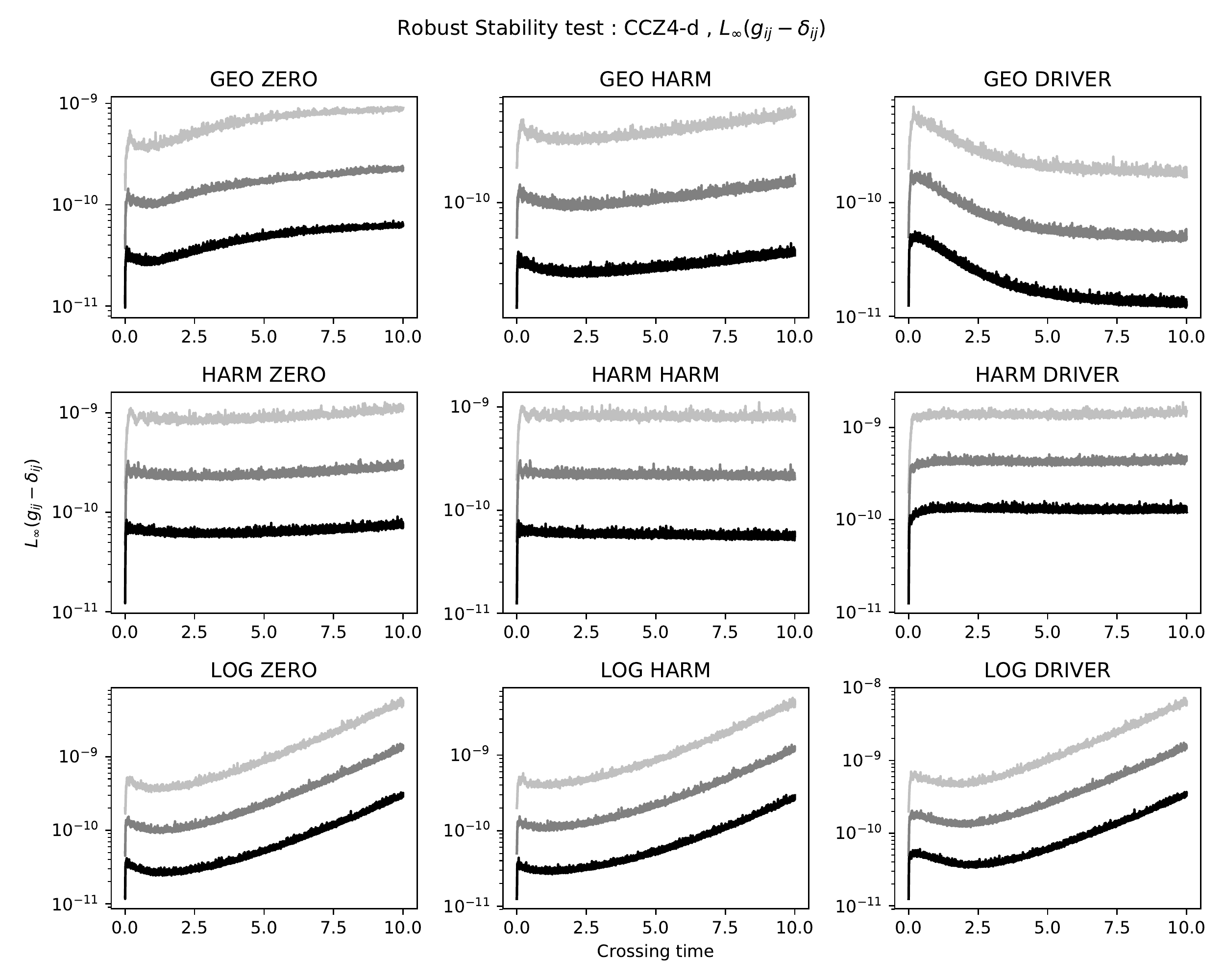} 
\caption{The $L_{\infty}$ norm of $g_{ij} - \de_{ij}$ as a function of the number of crossing times in the robust stability test of the CCZ4-d scheme for all nine gauges choices. Each plot contains the three resolutions $\ro \in \{ 1,2,4 \}$ corresponding to the light-gray, dark-gray and black lines, respectively.}
\label{fig:Rob_ga_CCZ4d}
\end{figure}

\begin{figure}
\includegraphics[width=\columnwidth]{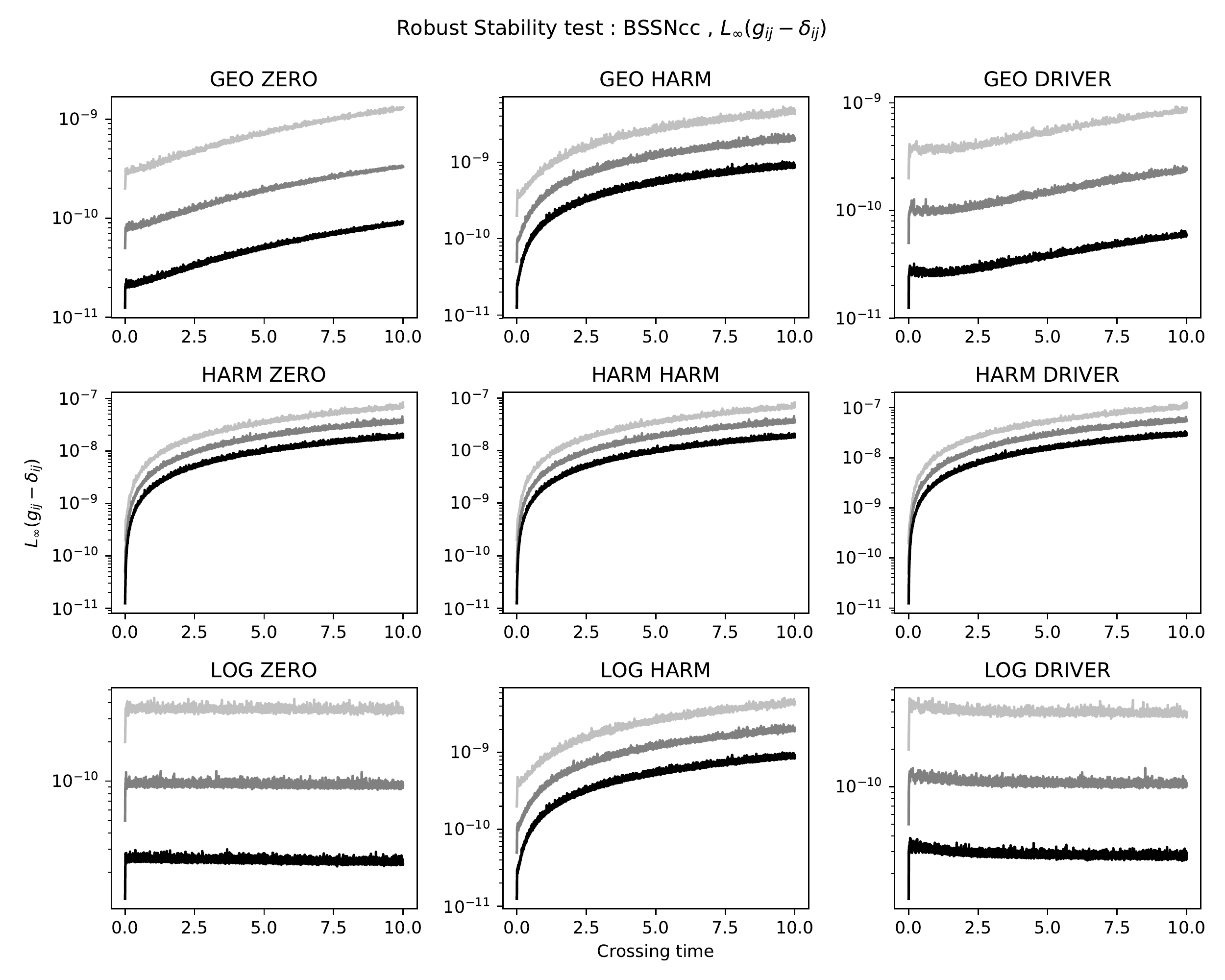} 
\caption{The $L_{\infty}$ norm of $g_{ij} - \de_{ij}$ as a function of the number of crossing times in the robust stability test of the BSSNcc scheme for all nine gauges choices. Each plot contains the three resolutions $\ro \in \{ 1,2,4 \}$ corresponding to the light-gray, dark-gray and black lines, respectively.}
\label{fig:Rob_ga_BSSNcc}
\end{figure}

\begin{figure}
\includegraphics[width=\columnwidth]{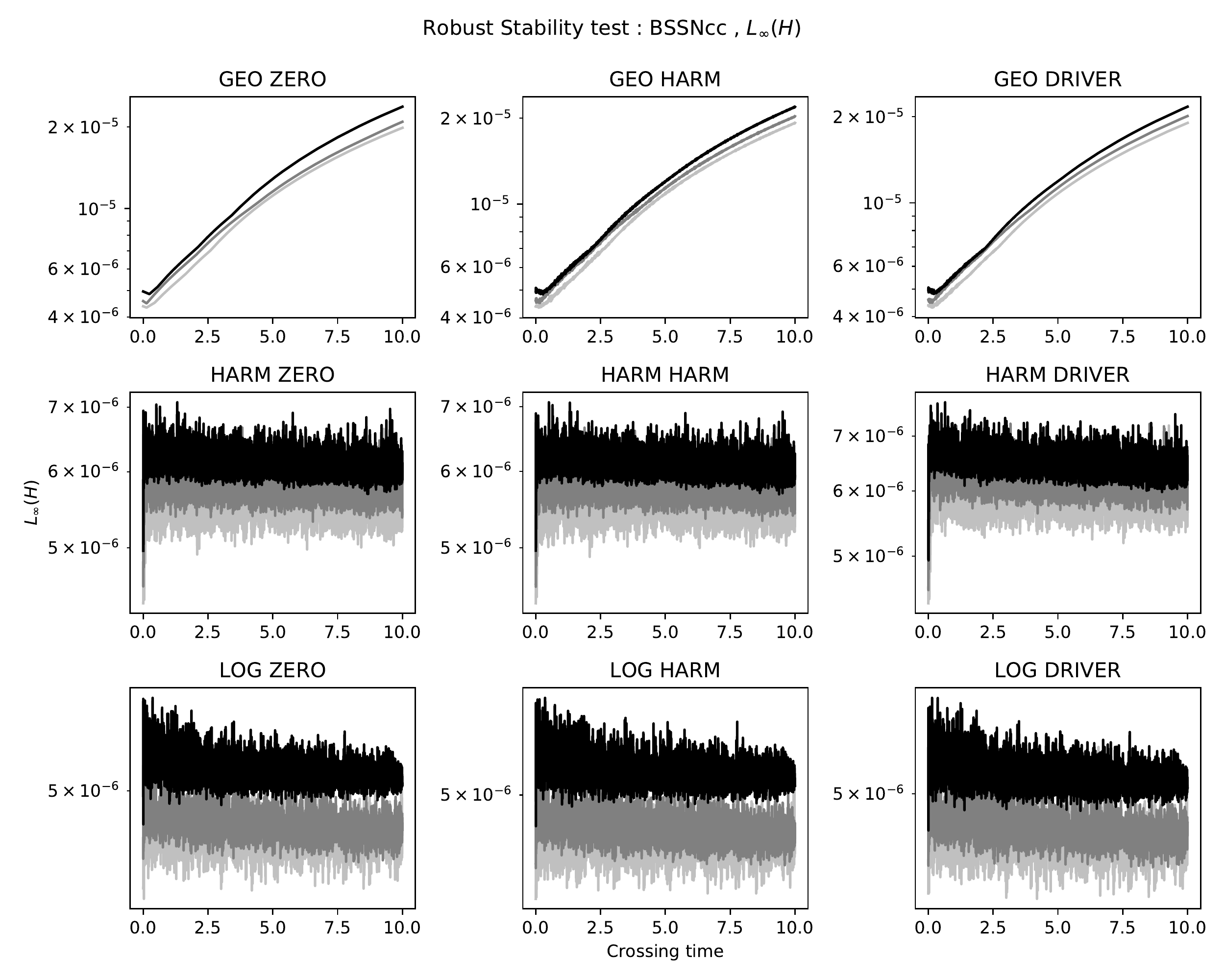} 
\caption{The $L_{\infty}$ norm of $H$ as a function of the number of crossing times in the robust stability test of the BSSNcc scheme for all nine gauges choices. Each plot contains the three resolutions $\ro \in \{ 1,2,4 \}$ corresponding to the light-gray, dark-gray and black lines, respectively.}
\label{fig:Rob_H_BSSNcc}
\end{figure}

\begin{itemize}

\item
The tests seem to pass for all schemes and gauges, in the sense that we observe the expected convergence of the $g_{ij} - \de_{ij}$ errors as $\ro$ grows.

\item
The best performance is clearly the one of the undamped Z4 schemes (CCZ4-u, Z4cc-u and Z4c-u) in that the monitored errors are always bounded for all gauges, with the only exception being the GEO-ZERO gauge where $L_{\infty}(g_{ij} - \de_{ij})$ and $L_{\infty}(\ti{Z}^i)$ grow. This fact is illustrated in fig. \ref{fig:Rob_ga_CCZ4u}.

\item
As in fig. \ref{fig:Rob_ga_CCZ4d}, all damped Z4 schemes (CCZ4-d, Z4cc-d and Z4c-d) systematically exhibit growing modes in the much used LOG slicing for all shift choices and sometimes also with GEO slicing, but are generically under control with HARM slicing. 

\item
For the BSSN schemes (BSSNcc and BSSNc) the noise is bounded for all monitored quantities only in the LOG-ZERO and LOG-DRIVER gauges (see fig. \ref{fig:Rob_ga_BSSNcc}). The Hamiltonian constraint $H$ is well-behaved for HARM slicing as well (see fig. \ref{fig:Rob_H_BSSNcc}). However, this is really a particularity of $H$, as none of the other quantities remains bounded in this slicing. As for the momentum constraint $M_i$, it is also well-behaved in the GEO-ZERO and GEO-DRIVER cases.

\end{itemize}

\section{Linearized wave test}

\subsection{Specifications} \label{sec:linewavespec}

In terms of the integrated variables, the analytical solution is given by
\beq
\al, \chi = 1 \, , \hspace{1cm} K = 0 \, , \hspace{1cm} \hat{\Ga}^i = 0 \, , 
\eeq
\beq
\ti{\ga}_{ij} = \de_{ij} + \( \begin{array}{ccc} -1 & 0 & 0 \\ 0 & 1 & 0 \\ 0 & 0 & 0  \end{array} \) H_s \, , \hspace{1cm} \ti{A}_{ij} = \( \begin{array}{ccc} -1 & 0 & 0 \\ 0 & 1 & 0 \\ 0 & 0 & 0  \end{array} \)  H_c \, ,
\eeq
where
\beq
H_s := A \sin \[ 2 \pi \( z - t \) \] \, , \hspace{1cm} H_c := A \pi \cos \[ 2 \pi \( z - t \) \] \, .
\eeq
This corresponds to the ``aligned" version, i.e. when the wave propagates in a grid direction, so that the numerical solution is symmetric in the two transverse directions. The ``diagonal" version is obtained by performing a 45 degree rotation in the $y$ direction, thus effectively probing a two-dimensional grid. One must then also rescale the wave-length by a factor of $\sqrt{2}$ in order to maintain the grid periodicity. The corresponding solution reads
\beq
\al, \chi = 1 \, , \hspace{1cm} K = 0 \, , \hspace{1cm} \hat{\Ga}^i = 0 \, , 
\eeq
\beq
\ti{\ga}_{ij} = \de_{ij} + \( \begin{array}{ccc} -1  & 0  & 0 \\ 0  & 1/2 & 1/2 \\ 0 & 1/2 & 1/2  \end{array} \) H_s \, , \hspace{1cm} \ti{A}_{ij} = \sqrt{2} \( \begin{array}{ccc} -1  & 0  & 0 \\ 0  & 1/2 & 1/2 \\ 0 & 1/2 & 1/2  \end{array} \) H_c \, ,
\eeq
where now
\beq
H_s := A \sin \[ 2 \pi \( z - y - \sqrt{2}\,t \) \] \, , \hspace{1cm} H_c := A \pi \cos \[ 2 \pi \( z - y - \sqrt{2}\,t \) \]  \, .
\eeq
The parameter choice here is
\beq
T = 1000 \, , \hspace{1cm}  C = 0.25  \, , \hspace{1cm} A = 10^{-8} \, , \hspace{1cm} \si = 0.8 \, .
\eeq
We find that the diagonal version provides qualitatively similar results as the aligned one, with the only difference being a larger error of at most one order of magnitude. Since this test corresponds to a large number of plots, we choose to display only the ones of the aligned version. For all nine gauge choices, we output  the evolution of the $L_{\infty}$ norm of $\De g_{xx}$, $\Te$, $\ti{Z}^i$, $H$ and $M_i$. 

On top of this, we also monitor the wave profile of $g_{xx} - 1$. However, instead of displaying the profiles themselves, as suggested in \cite{A2A1, A2A2}, we consider a different measure that provides more transparent information in our opinion, namely, the discrete Fourier transform in the $z$ direction at each crossing time. Denoting the integer momentum variable by $k \in \mathbb{Z}$, in the aligned case we have for instance
\beq
F_k(t) := \frac{1}{N} \sum_{n = 1}^N \big( g_{xx}(t,n \De x) - 1 \big) \exp \[ -2\pi i k n \De x \] \, ,
\eeq
so that the analytical solution corresponds to $F_k(t) = F \de_{k,1}$. We then choose to record
\begin{itemize}
\item
the relative phase to the analytical solution ${\rm arg} \( F_1 \)$,
\item
the relative error on the amplitude $\de F_1 := \( |F_1| - A \)/A$,
\item
the offset of the wave relatively to the amplitude $\de F_0 := |F_0|/A$,
\end{itemize}
at each crossing time and for all nine gauge choices (figs. \ref{fig:Linear_CCZ4-u_gxx_pao} to \ref{fig:Linear_BSSNc_gxx_pao}). Moreover, we consider the modulus of the full spectrum $|F_k|$ at the final time for the highest resolution $\ro = 4$ (figs. \ref{fig:Linear_CCZ4-u_gxx_spectra} and \ref{fig:Linear_BSSNc_gxx_spectra}) again for all nine gauge choices. For comparison, we also plot the numerical discrete Fourier transform of the analytical solution evaluated on the lattice. By proceeding thusly, instead of considering the exact continuous spectrum, we obtain a natural minimal threshold for the spectrum noise.

\subsection{Comments}

\begin{figure}
\includegraphics[width=\columnwidth]{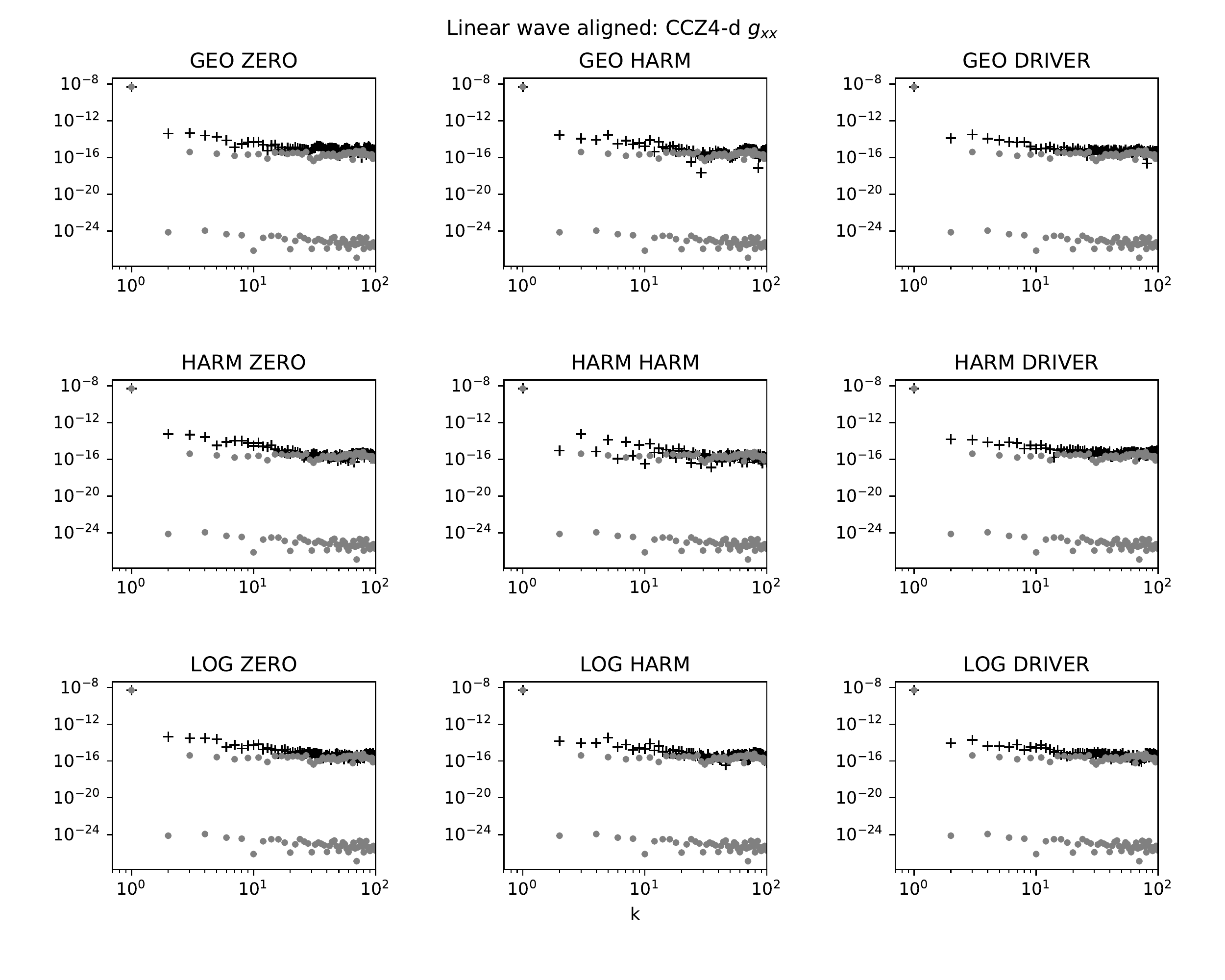} 
\caption{The modulus of the final spectra of $g_{xx} - 1$ in the aligned linear wave test of the CCZ4-d scheme at the highest resolution $\ro = 4$ and for all nine gauge choices. The black crosses correspond to the numerical solution, while the gray circles correspond to the analytical solution.}
\label{fig:Linear_CCZ4-d_gxx_spectra}
\end{figure}

\begin{figure}
\includegraphics[width=0.8\columnwidth]{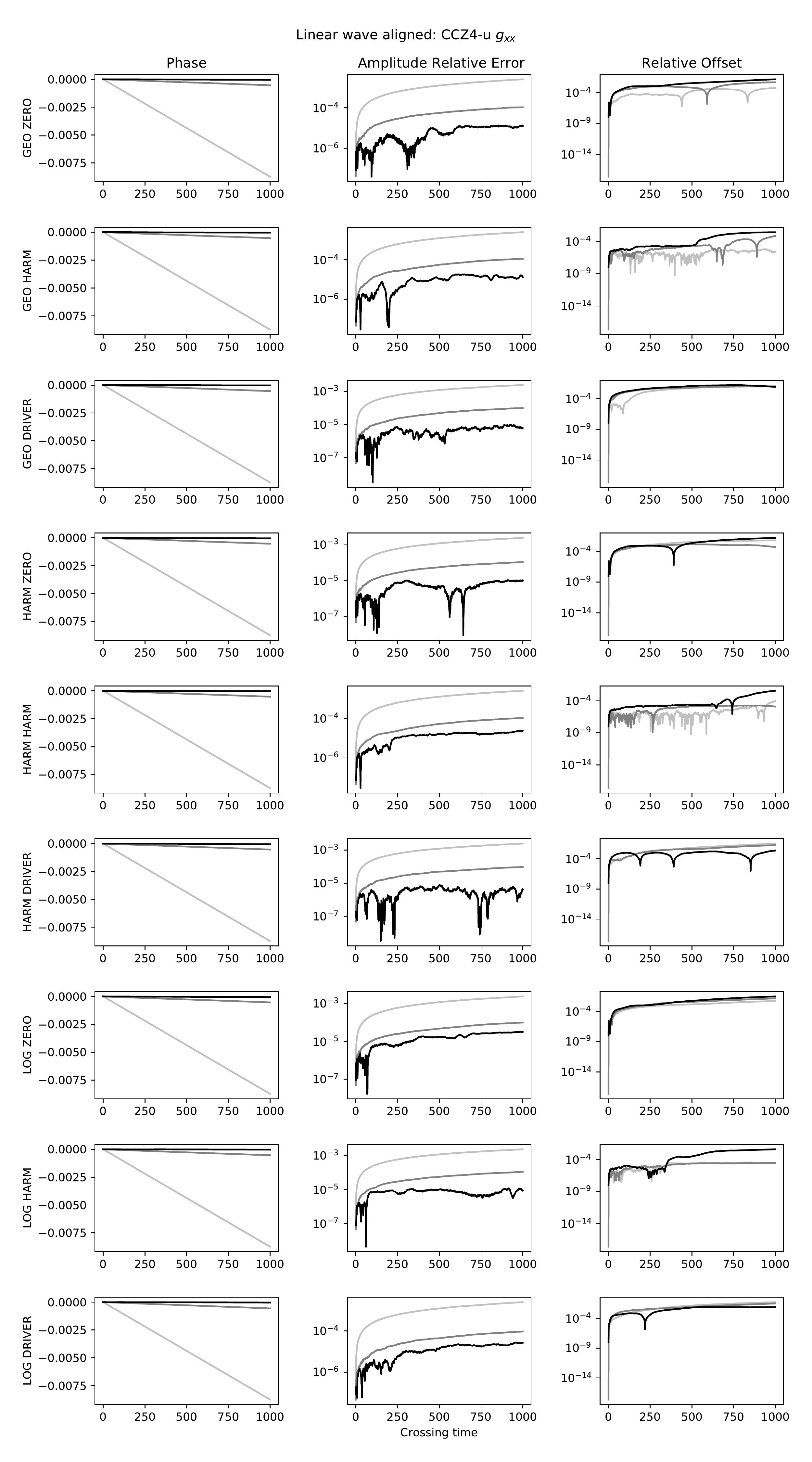} 
\caption{The relative phase ${\rm arg}\( F_1 \)$, relative error on the amplitude $\de F_1$ (``amplitude relative error") and the relative zero mode $\de F_0$  (``offset") of $g_{xx} - 1$ as a function of the number of crossing times in the aligned linear wave test of the CCZ4-u scheme in all nine gauge choices. Each plot contains the three resolutions $\ro \in \{ 1,2,4 \}$ corresponding to the light-gray, dark-gray and black lines, respectively.}
\label{fig:Linear_CCZ4-u_gxx_pao}
\end{figure}

\begin{figure}
\includegraphics[width=0.8\columnwidth]{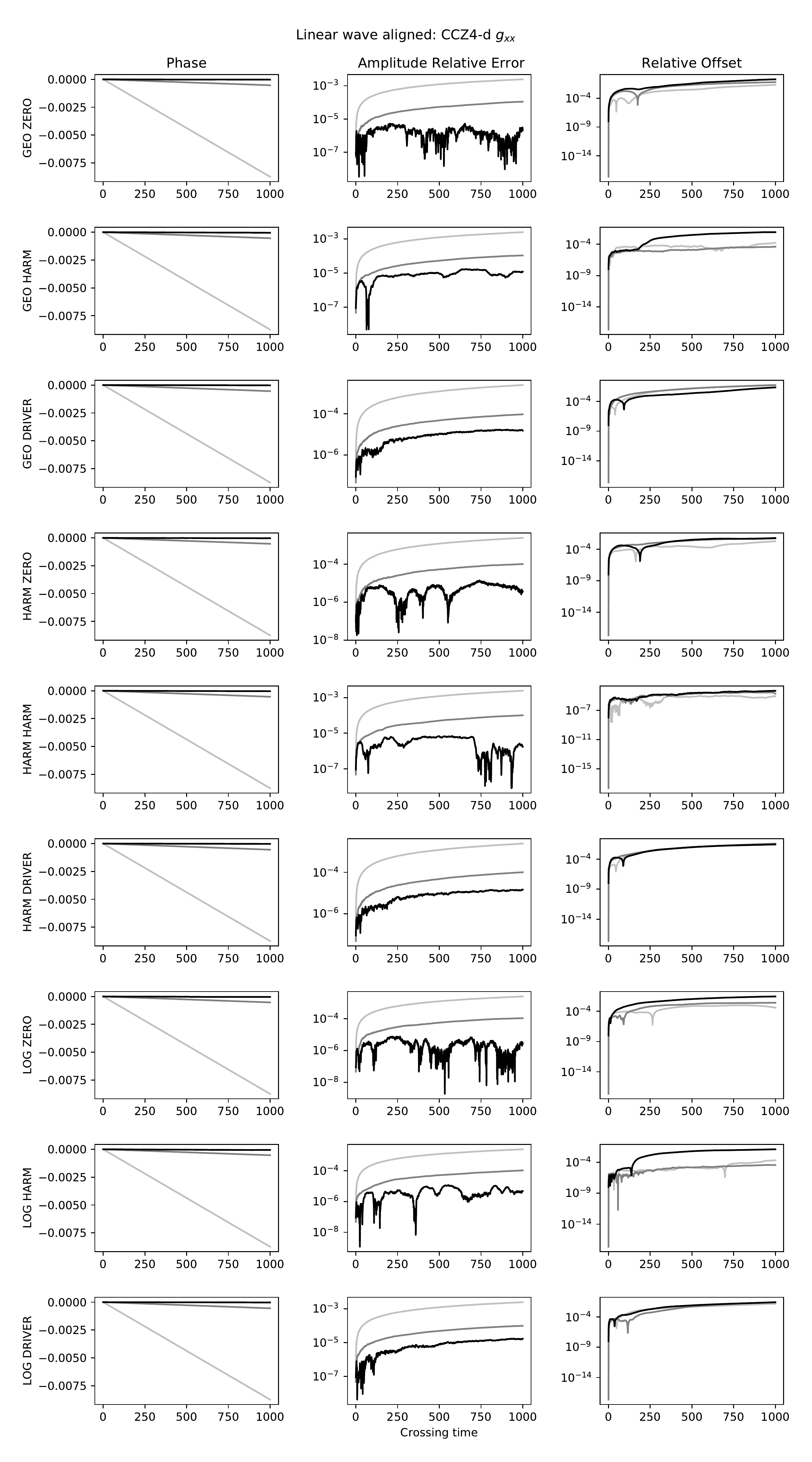} 
\caption{The relative phase ${\rm arg}\( F_1 \)$, relative error on the amplitude $\de F_1$ (``amplitude relative error") and the relative zero mode $\de F_0$  (``offset") of $g_{xx} - 1$ as a function of the number of crossing times in the aligned linear wave test of the CCZ4-d scheme in all nine gauge choices. Each plot contains the three resolutions $\ro \in \{ 1,2,4 \}$ corresponding to the light-gray, dark-gray and black lines, respectively.}
\label{fig:Linear_CCZ4-d_gxx_pao}
\end{figure}

\begin{figure}
\includegraphics[width=0.8\columnwidth]{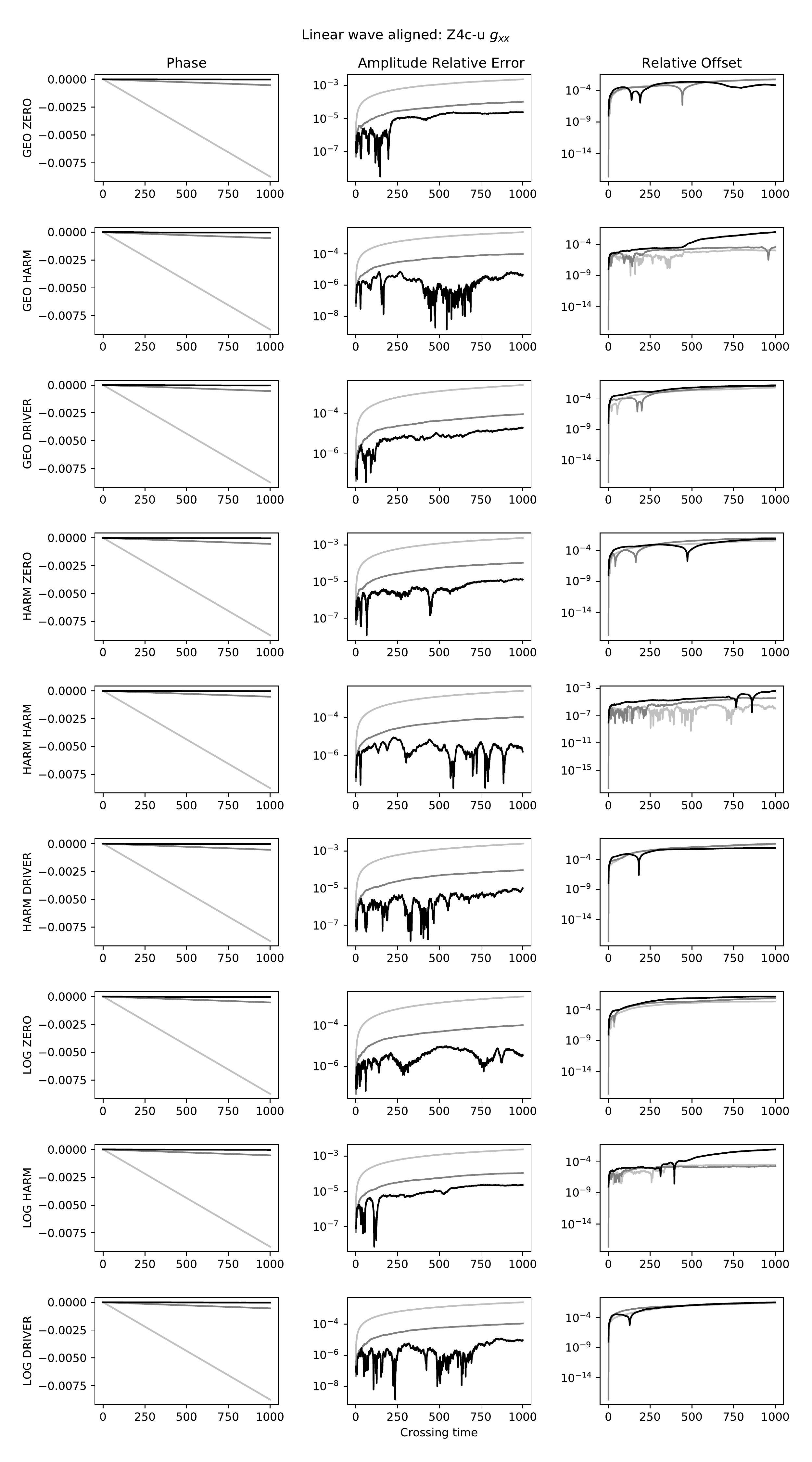} 
\caption{The relative phase ${\rm arg}\( F_1 \)$, relative error on the amplitude $\de F_1$ (``amplitude relative error") and the relative zero mode $\de F_0$  (``offset") of $g_{xx} - 1$ as a function of the number of crossing times in the aligned linear wave test of the Z4c-u scheme in all nine gauge choices. Each plot contains the three resolutions $\ro \in \{ 1,2,4 \}$ corresponding to the light-gray, dark-gray and black lines, respectively.}
\label{fig:Linear_Z4c-u_gxx_pao}
\end{figure}

\begin{figure}
\includegraphics[width=0.8\columnwidth]{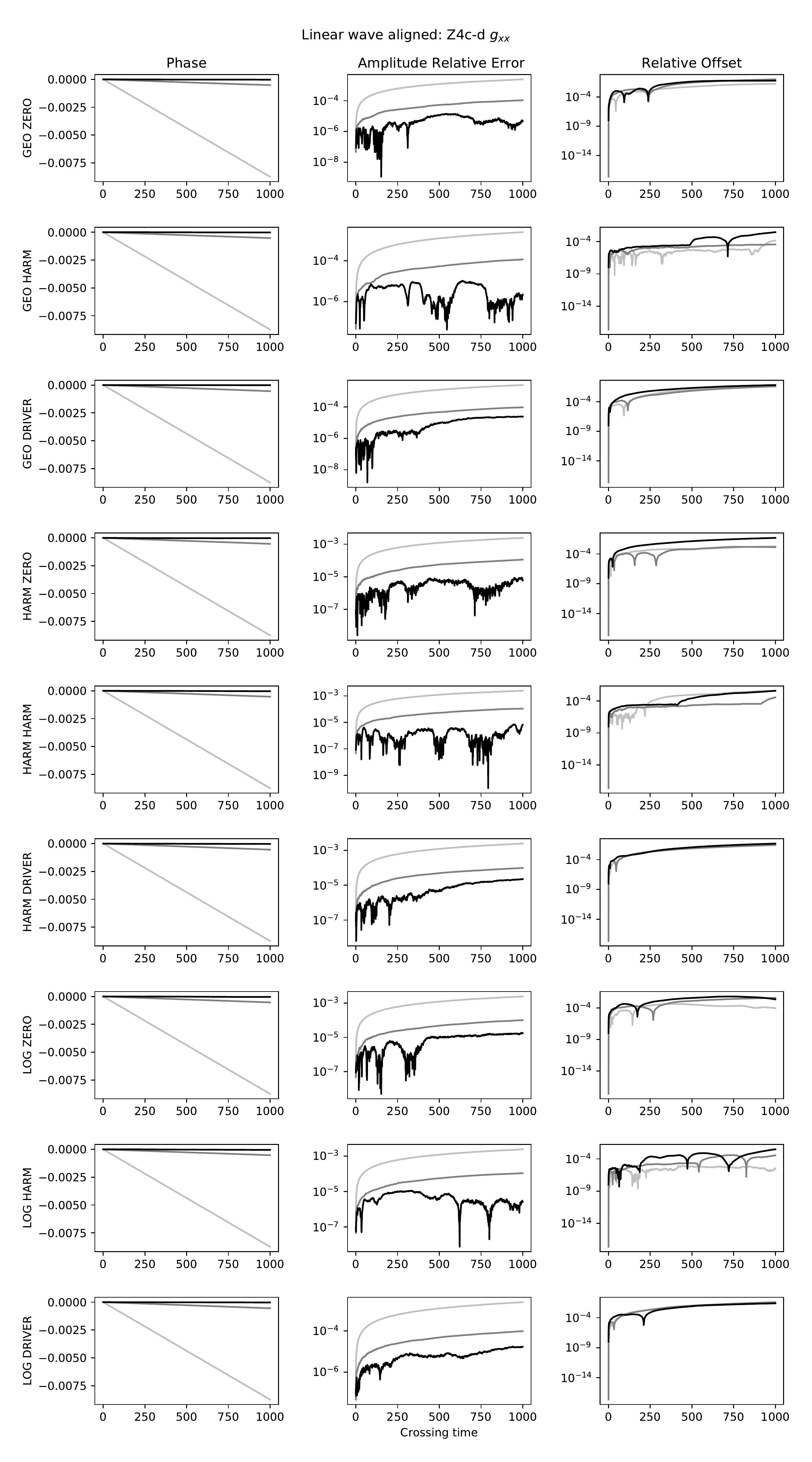} 
\caption{The relative phase ${\rm arg}\( F_1 \)$, relative error on the amplitude $\de F_1$ (``amplitude relative error") and the relative zero mode $\de F_0$  (``offset") of $g_{xx} - 1$ as a function of the number of crossing times in the aligned linear wave test of the Z4c-d scheme in all nine gauge choices. Each plot contains the three resolutions $\ro \in \{ 1,2,4 \}$ corresponding to the light-gray, dark-gray and black lines, respectively.}
\label{fig:Linear_Z4c-d_gxx_pao}
\end{figure}

\begin{figure}
\includegraphics[width=0.8\columnwidth]{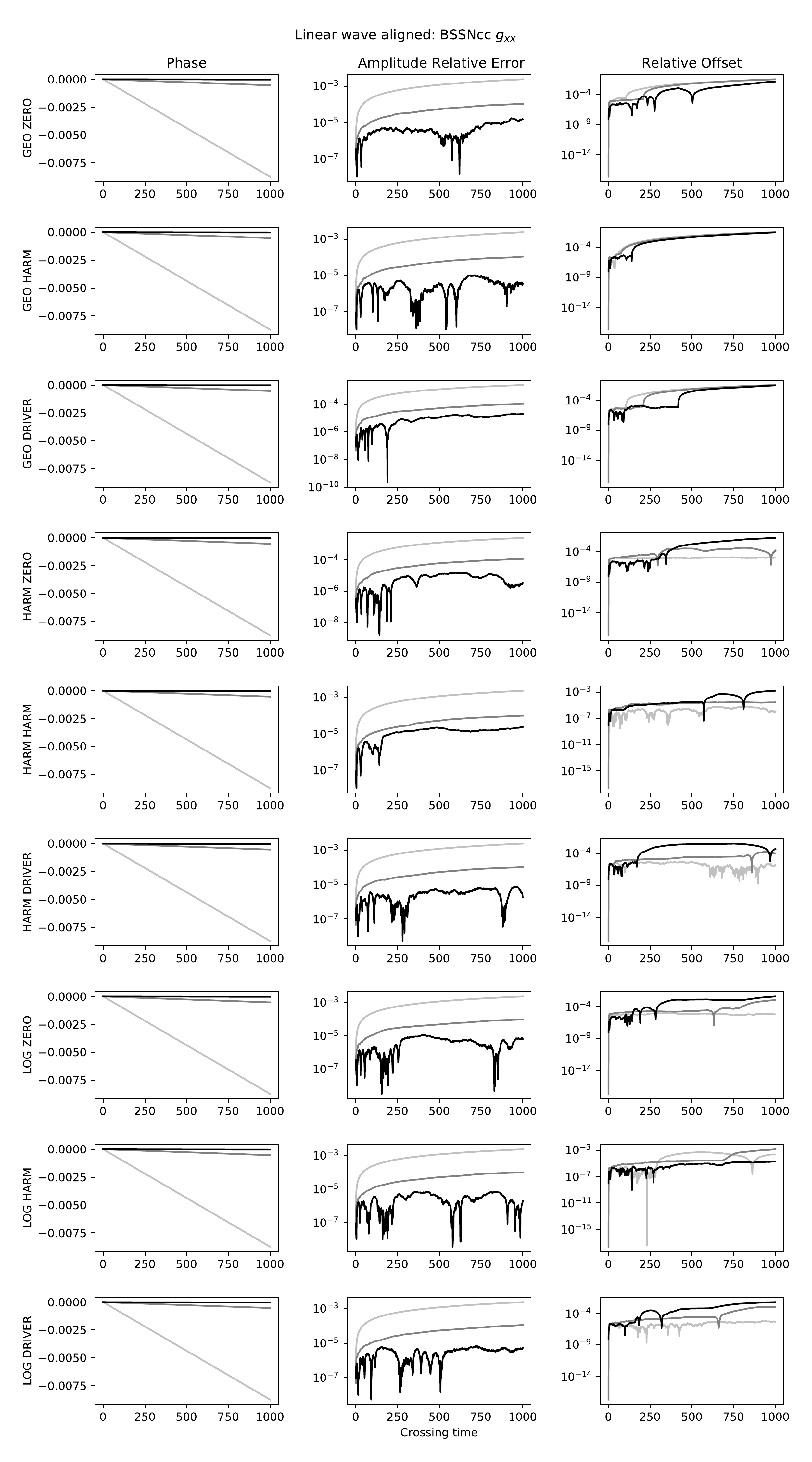} 
\caption{The relative phase ${\rm arg}\( F_1 \)$, relative error on the amplitude $\de F_1$ (``amplitude relative error") and the relative zero mode $\de F_0$  (``offset") of $g_{xx} - 1$ as a function of the number of crossing times in the aligned linear wave test of the BSSNcc scheme in all nine gauge choices. Each plot contains the three resolutions $\ro \in \{ 1,2,4 \}$ corresponding to the light-gray, dark-gray and black lines, respectively.}
\label{fig:Linear_BSSNcc_gxx_pao}
\end{figure}

\begin{figure}
\includegraphics[width=0.8\columnwidth]{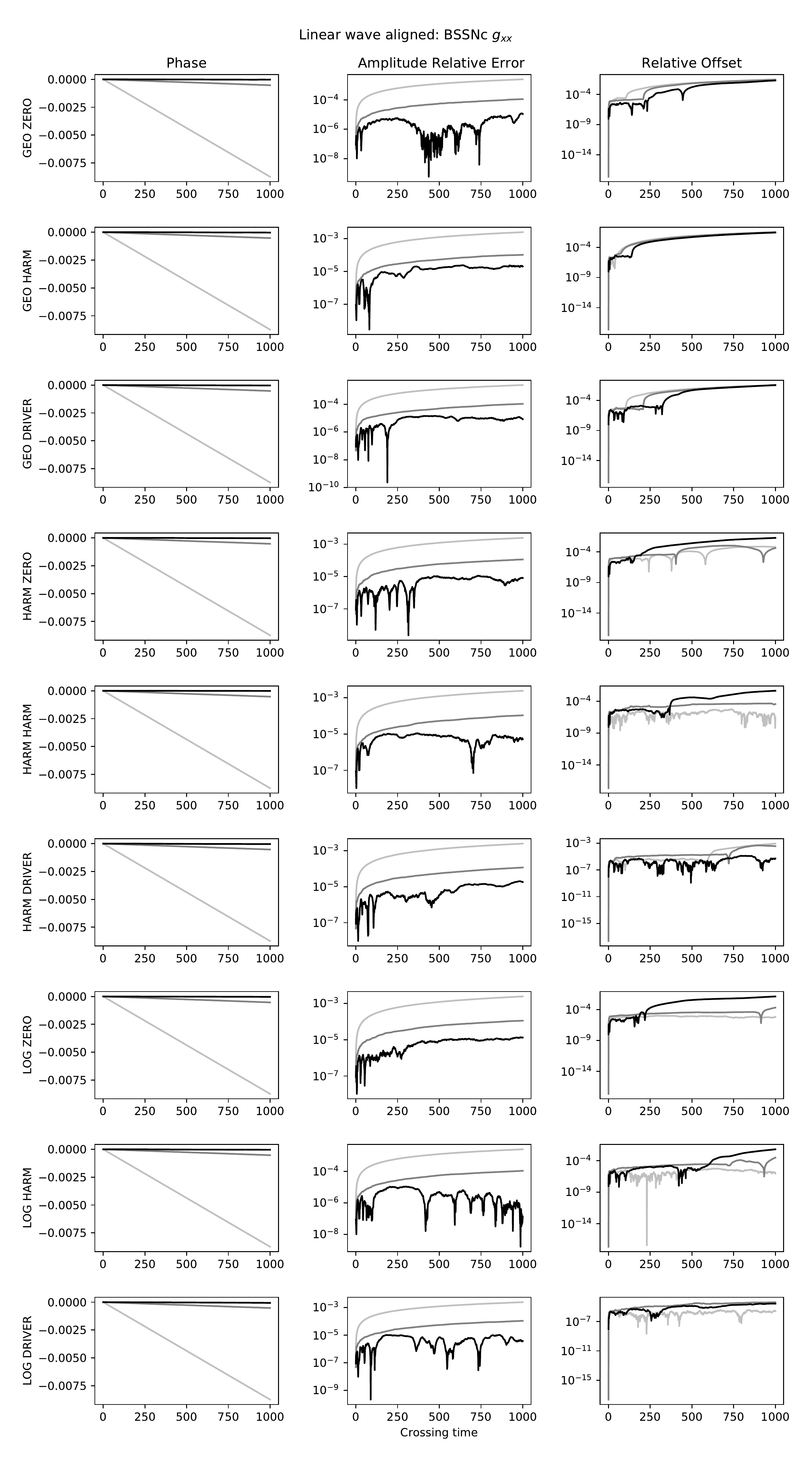} 
\caption{The relative phase ${\rm arg}\( F_1 \)$, relative error on the amplitude $\de F_1$ (``amplitude relative error") and the relative zero mode $\de F_0$  (``offset") of $g_{xx} - 1$ as a function of the number of crossing times in the aligned linear wave test of the BSSNc scheme in all nine gauge choices. Each plot contains the three resolutions $\ro \in \{ 1,2,4 \}$ corresponding to the light-gray, dark-gray and black lines, respectively.}
\label{fig:Linear_BSSNc_gxx_pao}
\end{figure}

\begin{figure}
\includegraphics[width=\columnwidth]{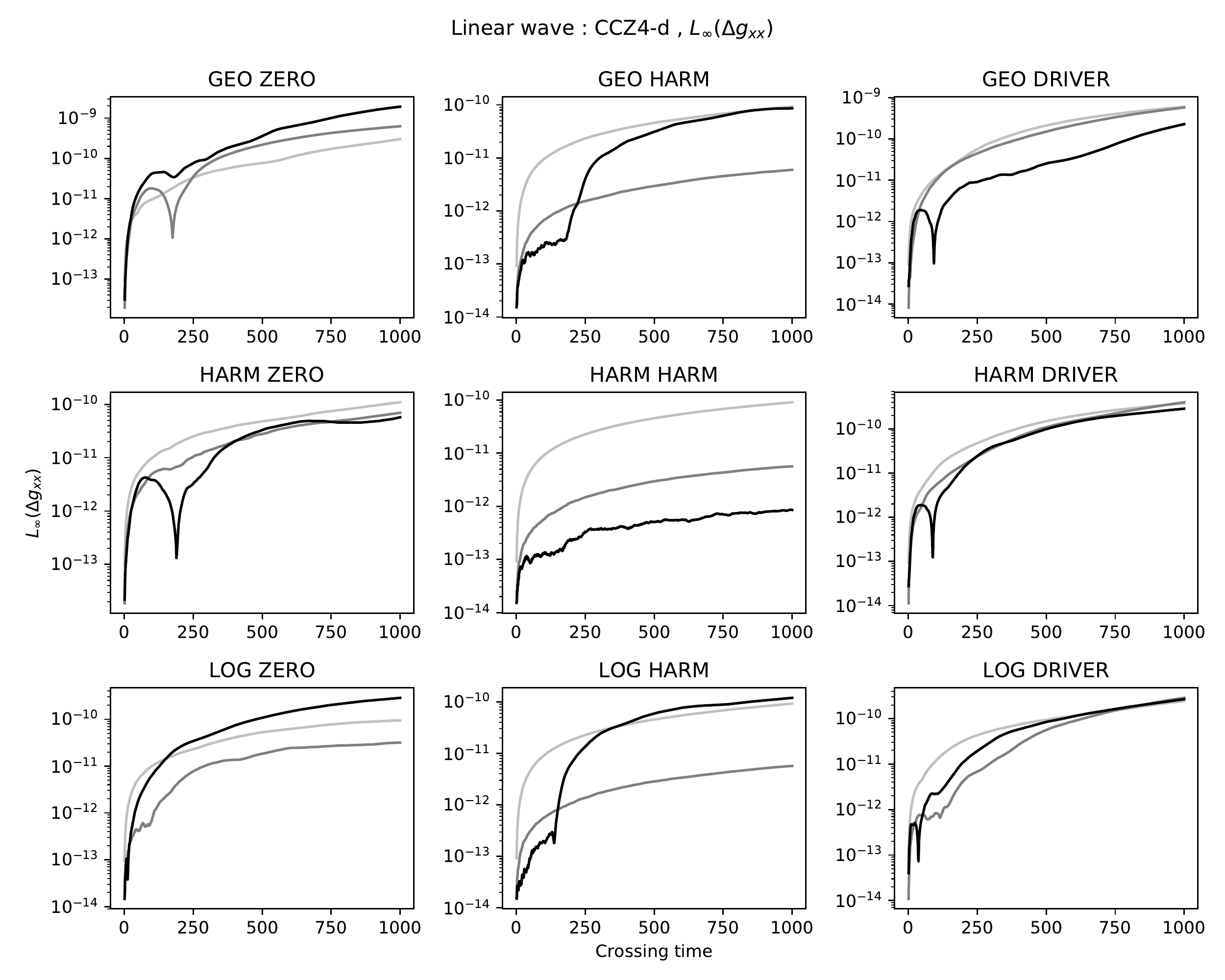} 
\caption{The $L_{\infty}$ norm of the absolute error $\De g_{xx}$ as a function of the number of crossing times for the aligned linear wave test of the CCZ4-d scheme in all nine gauge choices. Each plot contains the three resolutions $\ro \in \{ 1,2,4 \}$ corresponding to the light-gray, dark-gray and black lines, respectively.}
\label{fig:Linear_CCZ4-d_Linf_error_gxx}
\end{figure}

\begin{figure}
\includegraphics[width=\columnwidth]{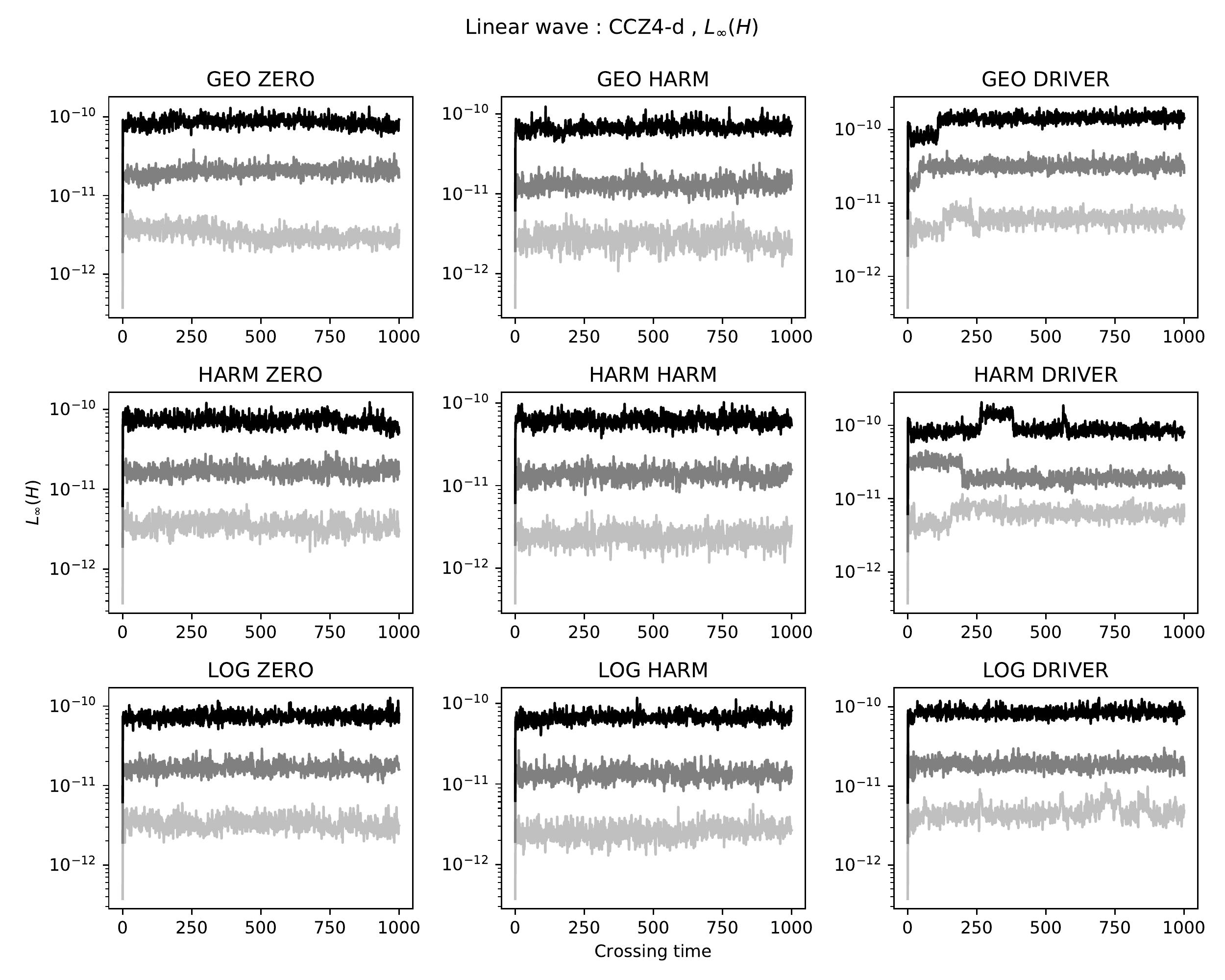} 
\caption{The $L_{\infty}$ norm of $H$ as a function of the number of crossing times for the aligned linear wave test of the CCZ4-d scheme in all nine gauge choices. Each plot contains the three resolutions $\ro \in \{ 1,2,4 \}$ corresponding to the light-gray, dark-gray and black lines, respectively.}
\label{fig:Linear_CCZ4-d_Linf_H}
\end{figure}

\begin{itemize}

\item
From the Fourier mode analysis we see that the dominating error lies in the wave's offset, i.e. its $k = 0$ mode. Indeed, in the final spectra (see e.g. fig. \ref{fig:Linear_CCZ4-d_gxx_spectra}) we see that $|F_1|$ fits remarkably well the analytical solution, with the $k > 1$ modes arising only several orders of magnitude below. This is also confirmed more quantitatively by looking at the phases and amplitudes in figs. \ref{fig:Linear_CCZ4-u_gxx_pao} to \ref{fig:Linear_BSSNc_gxx_pao}. We generically have $\de F_1 \sim 10^{-6}, 10^{-5}$ in the highest resolution $\ro = 4$ (black lines) and $\de F_1 \sim 10^{-3}$ in the lowest resolution $\ro = 1$ (light-gray lines). On the other hand, $\de F_0$ is generically larger by several orders of magnitude. Interestingly, this error is even more pronounced for the highest resolution $\ro = 4$ in general.

\item
The CCZ4-d scheme in HARM-HARM gauge, the BSSNcc in LOG-HARM gauge and BSSNc in HARM-DRIVER and LOG-DRIVER gauge yield the smallest $\De g_{xx}$ error generically. Given the previous point, this is directly obvious by looking at the behavior of the offset in figs. \ref{fig:Linear_CCZ4-u_gxx_pao} to \ref{fig:Linear_BSSNc_gxx_pao}, which is several orders of magnitude lower than in other gauges.

\item 
The $L_{\infty}$ norm of $\De g_{xx}$, although decreasing with resolution generically, never seems to converge (see for example fig. \ref{fig:Linear_CCZ4-d_Linf_error_gxx}). 

\item
The constraint violation systematically diverges with resolution increase for all schemes and gauges. We have also considered the $L_2$ norms (not displayed) and the behavior is similar (see for example fig. \ref{fig:Linear_CCZ4-d_Linf_H}).

\item 
Damping $\ka > 0$ stabilizes constraint violation in Z4 schemes, except for the Z4c scheme in the case of the $\ti{Z}^i$ constraint. This constraint also diverges in BSSN schemes.

\end{itemize}

\section{Gauge wave test}

\subsection{Specifications}

This test is defined for harmonic slicing and zero shift (HARM-ZERO). As in the case of the linear wave test, we consider the aligned solution
\beq
\al = \sqrt{1 - H_s} \, , \hspace{1cm} \chi = \( 1 - H_s \)^{-1/3} \, , \hspace{1cm} K = - \frac{H_c}{\( 1 - H_s \)^{3/2}} \, , 
\eeq
\beq
\hat{\Ga}^{1,2} = 0 \, , \hspace{1cm} \hat{\Ga}^3 = -\frac{4}{3} \frac{H_c}{\( 1 - H_s \)^{5/3}} \, , 
\eeq
\beq
\ti{\ga}_{ij} = \( \begin{array}{ccc} \( 1 - H_s \)^{-1/3}  & 0  & 0 \\ 0  & \( 1 - H_s \)^{-1/3}  & 0 \\ 0 & 0 & \( 1 - H_s \)^{2/3}  \end{array} \)  \, , 
\eeq
\beq
\ti{A}_{ij} = \frac{1}{3}\, H_c \( \begin{array}{ccc} \( 1 - H_s \)^{-11/6} & 0 & 0 \\ 0 & \( 1 - H_s \)^{-11/6} & 0 \\ 0 & 0 & -2 \( 1 - H_s \)^{-5/6} \end{array} \) \, ,
\eeq
where
\beq
H_s := A \sin \[ 2 \pi \( z - t \) \] \, , \hspace{1cm} H_c := A \pi \cos \[ 2 \pi \( z - t \) \]  \, ,
\eeq
and the diagonal solution
\beq
\al = \sqrt{1 - H_s} \, , \hspace{1cm} \chi = \( 1 - H_s \)^{-1/3} \, , \hspace{1cm} K = - \frac{\sqrt{2} \, H_c}{\( 1 - H_s \)^{3/2}} \, , 
\eeq
\beq
\hat{\Ga}^1 = 0 \, , \hspace{1cm} \hat{\Ga}^2 = \frac{4}{3} \frac{H_c}{\( 1 - H_s \)^{5/3}} \, , \hspace{1cm} \hat{\Ga}^3 = -\frac{4}{3} \frac{H_c}{\( 1 - H_s \)^{5/3}} \, , 
\eeq
\beq
\ti{\ga}_{ij} = \frac{1}{\( 1 - H_s \)^{1/3}} \( \begin{array}{ccc} 1  & 0  & 0 \\ 0  & 1 - \frac{1}{2}\, H_s  & \frac{1}{2}\, H_s \\ 0 & \frac{1}{2}\, H_s & 1 - \frac{1}{2}\, H_s  \end{array} \)  \, , 
\eeq
\beq
\ti{A}_{ij} = \frac{\sqrt{2}}{3} \frac{H_c}{\( 1 - H_s \)^{11/6}} \( \begin{array}{ccc} 1 & 0 & 0 \\ 0 & -\frac{1}{2} \( 1 - 2 H_s \) & \frac{1}{2} \( 3 - 2 H_s \) \\ 0 & \frac{1}{2}  \( 3 - 2 H_s \) & -\frac{1}{2} \( 1 - 2 H_s \)  \end{array} \) \, ,
\eeq
where now
\beq
H_s := A \sin \[ 2 \pi \( z - y - \sqrt{2}\,t \) \] \, , \hspace{1cm} H_c := A \pi \cos \[ 2 \pi \( z - y - \sqrt{2}\,t \) \]  \, .
\eeq
The parameter choice is 
\beq
T = 1000 \, , \hspace{1cm}  C = 0.25  \, , \hspace{1cm} A = 0.1 \, , \hspace{1cm} \si = 1 \, .
\eeq
For both the aligned and diagonal versions, we output the $L_{\infty}$ norm of the absolute errors $\De g_{tt}$, of $\De g_{zz}$ and of the constraints $\Te$, $\ti{Z}^i$, $H$ and $M_i$. 

\begin{figure}
\includegraphics[width=\columnwidth]{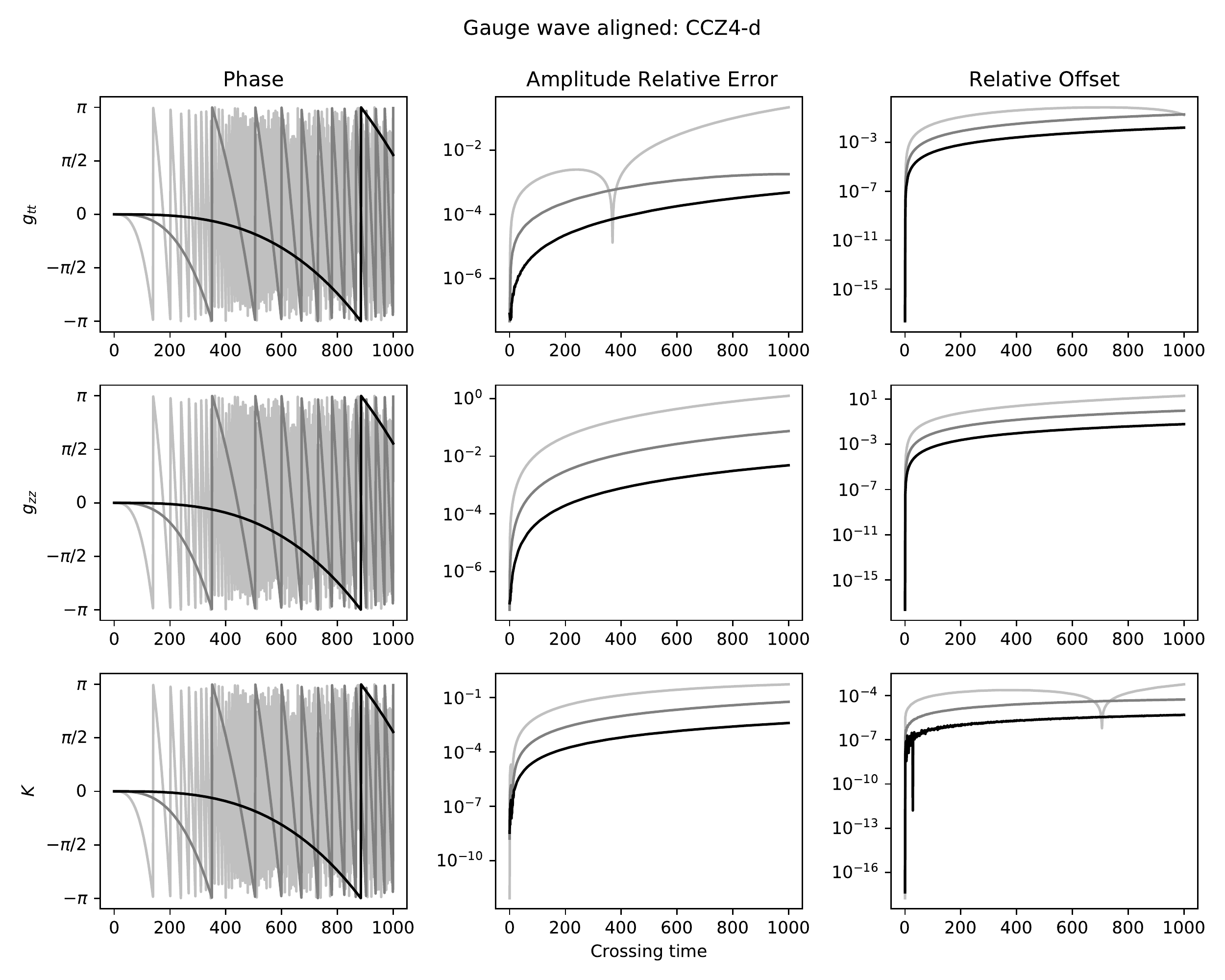} 
\caption{The relative phase ${\rm arg}\( F_1 \)$, relative error on the amplitude $\de F_1$ (``amplitude relative error") and the relative zero mode $\de F_0$ (``offset") of $g_{tt} + 1$, $g_{zz} - 1$ and $K$ as a function of the number of crossing times in the aligned gauge wave test for the CCZ4-d scheme. Each plot contains the three resolutions $\ro \in \{ 1,2,4 \}$ corresponding to the light-gray, dark-gray and black lines, respectively.}
\label{fig:Gauge_CCZ4d_pao}
\end{figure}

\begin{figure}
\includegraphics[width=\columnwidth]{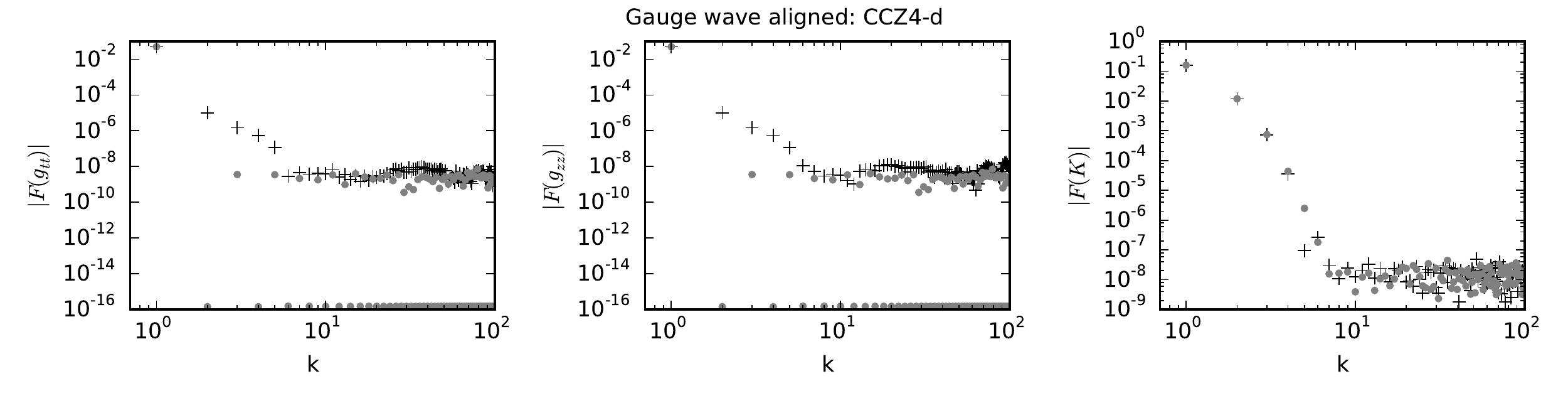} 
\caption{The modulus of the final spectra of $g_{tt} + 1$, $g_{zz} - 1$ and $K$ in the aligned gauge wave test for the CCZ4-d scheme at the highest resolution $\ro = 4$. The black crosses correspond to the numerical solution, while the gray circles correspond to the analytical solution.}
\label{fig:Gauge_cc_Z4_pao}
\end{figure}

For the scheme that does not crash (CCZ4-d), we also monitor the spectrum of $g_{tt} + 1$, $g_{zz} - 1$ and $K$ for the aligned version (see section \ref{sec:linewavespec} for a discussion of this measure). The quantities ${\rm arg}\( F_1\)$, $\de F_1$ and $\de F_0$ of the linearized wave test are defined here analogously for each aforementioned quantity and relatively to its typical analytical amplitude, and are displayed in fig. \ref{fig:Gauge_CCZ4d_pao}. The modulus of the final spectrum for the highest resolution $\ro = 4$ is given in fig. \ref{fig:Gauge_cc_Z4_pao}.

\subsection{Comments}

\begin{figure}
\includegraphics[width=\columnwidth]{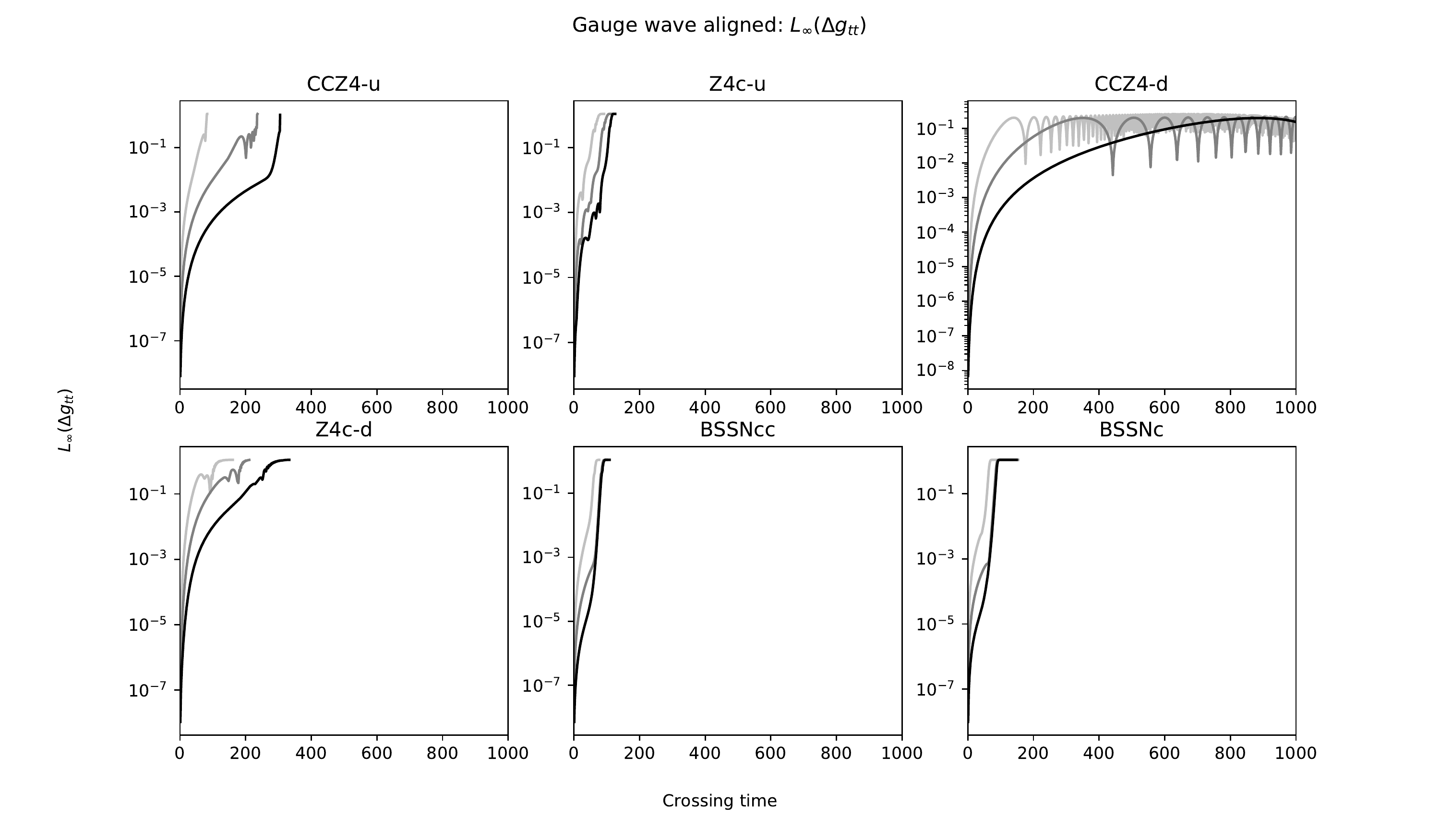} 
\caption{The $L_{\infty}$ norm of the absolute error $\De g_{tt}$ as a function of the number of crossing times in the aligned gauge wave test for all eight schemes. Each plot contains the three resolutions $\ro \in \{ 1,2,4 \}$ corresponding to the light-gray, dark-gray and black lines, respectively.}
\label{fig:Gauge_gtt_align}
\end{figure}

\begin{figure}
\includegraphics[width=\columnwidth]{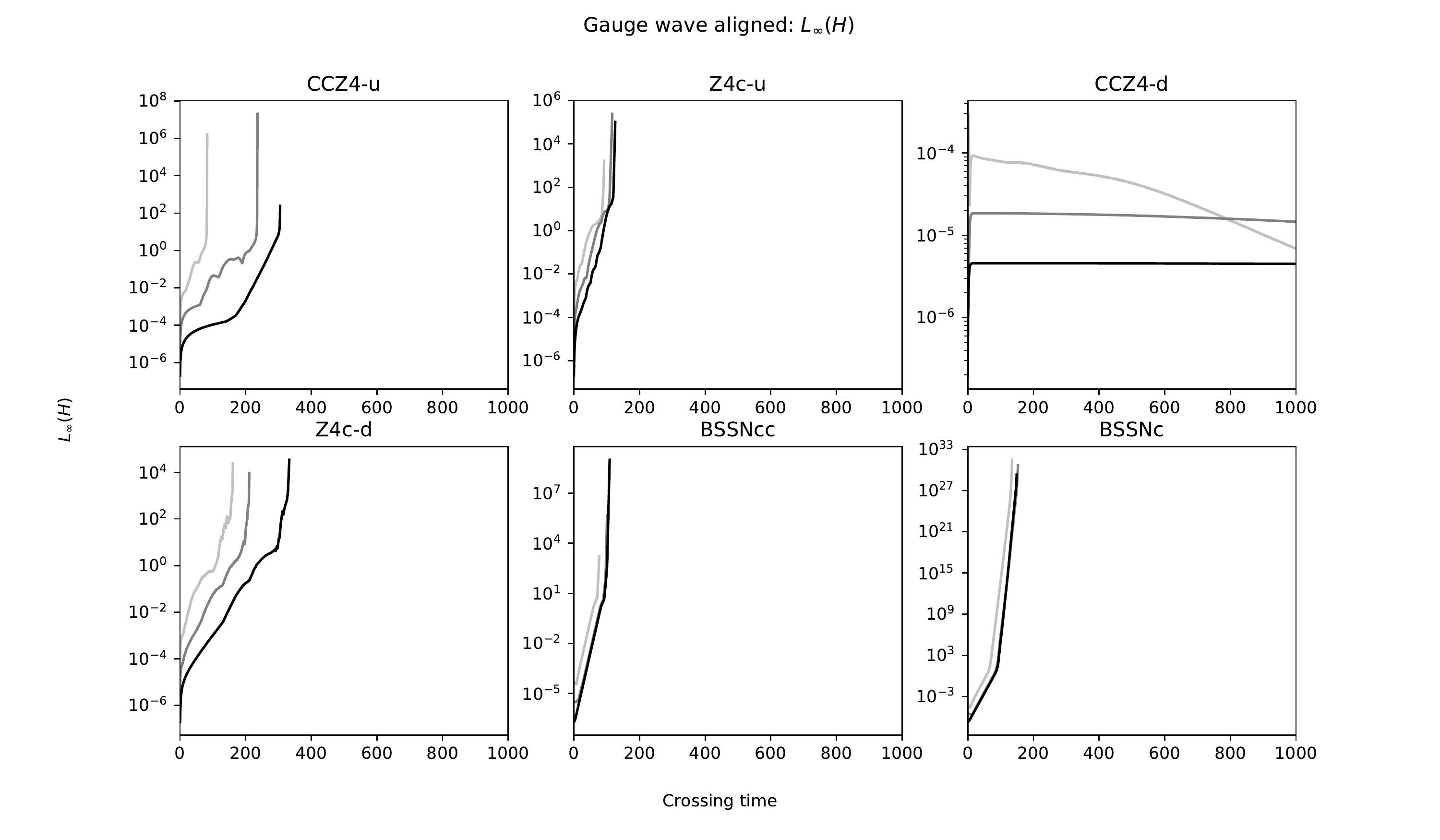} 
\caption{The $L_{\infty}$ norm of $H$ as a function of the number of crossing times in the aligned gauge wave test for all eight schemes. Each plot contains the three resolutions $\ro \in \{ 1,2,4 \}$ corresponding to the light-gray, dark-gray and black lines, respectively.}
\label{fig:Gauge_H_align}
\end{figure}

\begin{itemize}

\item
The only scheme that does not crash is CCZ4-d and yields very similar results for both the aligned and diagonal version. In that scheme, constraint violations are under control and the only difference is a slightly larger constraint violation in the diagonal case. The undamped Z4 schemes, as well as Z4c-d, are able to last a bit longer than the BSSN schemes before crashing. The rest of the comments concern the non-crashing cases. Illustrations of these facts are shown in fig. \ref{fig:Gauge_gtt_align} and fig. \ref{fig:Gauge_H_align}).

\item 
In fig. \ref{fig:Gauge_cc_Z4_pao} we have included the absolute value of the spectrum of $K$, which contains several modes above $k = 1$, and we see that we can accurately capture at least the first four. These span four orders of magnitude, so the accuracy at the level of the mode amplitudes is very good.

\item
From fig. \ref{fig:Gauge_CCZ4d_pao}, we see that the major error comes from the phase. Indeed, even with the highest resolution $\ro = 1$, the relative phase reaches almost a full $2\pi$ at $T = 1000$.

\end{itemize}

\section{Shifted gauge wave test}

\subsection{Specifications}

This test is defined for harmonic gauge (HARM-HARM). As in the previous two tests, we consider the aligned solution
\beq
\al = \( 1 + H_s \)^{-1/2} \, , \hspace{1cm} \chi = \( 1 + H_s \)^{-1/3} \, , \hspace{1cm} K = - \frac{H_c}{\( 1 + H_s \)^{3/2}} \, , 
\eeq
\beq
\hat{\Ga}^{1,2} = 0 \, , \hspace{1cm} \hat{\Ga}^3 = \frac{4}{3} \frac{H_c}{\( 1 + H_s \)^{5/3}} \, , \hspace{1cm} \be^{1,2} = 0 \, , \hspace{1cm} \be^3 = - \frac{H_s}{1 + H_s}  \, ,
\eeq
\beq
\ti{\ga}_{ij} = \( \begin{array}{ccc} \( 1 + H_s \)^{-1/3}  & 0  & 0 \\ 0  & \( 1 + H_s \)^{-1/3}  & 0 \\ 0 & 0 & \( 1 + H_s \)^{2/3}  \end{array} \)  \, , 
\eeq
\beq
\ti{A}_{ij} = \frac{1}{3}\, H_c \( \begin{array}{ccc} \( 1 + H_s \)^{-11/6} & 0 & 0 \\ 0 & \( 1 + H_s \)^{-11/6} & 0 \\ 0 & 0 & -2 \( 1 + H_s \)^{-5/6} \end{array} \) \, ,
\eeq
where
\beq
H_s := A \sin \[ 2 \pi \( z - t \) \] \, , \hspace{1cm} H_c := A \pi \cos \[ 2 \pi \( z - t \) \]  \, ,
\eeq
and the diagonal solution
\beq
\al = \( 1 + H_s \)^{-1/2} \, , \hspace{1cm} \chi = \( 1 + H_s \)^{-1/3} \, , \hspace{1cm} K = - \frac{\sqrt{2}\,H_c}{\( 1 + H_s \)^{3/2}}  \, , 
\eeq
\beq
\hat{\Ga}^1 = 0 \, , \hspace{1cm} \hat{\Ga}^2 = -\frac{4}{3} \frac{H_c}{\( 1 + H_s \)^{5/3}} \, , \hspace{1cm} \hat{\Ga}^3 = \frac{4}{3} \frac{H_c}{\( 1 + H_s \)^{5/3}} \, , 
\eeq
\beq
\be^1 = 0 \, , \hspace{1cm} \be^2 = \frac{H_s}{1 + H_s}   \, , \hspace{1cm} \be^3 = - \frac{H_s}{1 + H_s} \, , 
\eeq
\beq
\ti{\ga}_{ij} = \frac{1}{\( 1 + H_s \)^{1/3}} \( \begin{array}{ccc} 1  & 0  & 0 \\ 0  & 1 + \frac{1}{2}\, H_s  & -\frac{1}{2}\, H_s \\ 0 & -\frac{1}{2}\, H_s & 1 + \frac{1}{2}\, H_s  \end{array} \)  \, , 
\eeq
\beq
\ti{A}_{ij} = \frac{\sqrt{2}}{3} \frac{H_c}{\( 1 + H_s \)^{11/6}} \( \begin{array}{ccc} 1 & 0 & 0 \\ 0 & -\frac{1}{2} \( 1 + 2 H_s \) & \frac{1}{2} \( 3 + 2 H_s \) \\ 0 & \frac{1}{2}  \( 3 + 2 H_s \) & -\frac{1}{2} \( 1 + 2 H_s \)  \end{array} \) \, ,
\eeq
where now
\beq
H_s := A \sin \[ 2 \pi \( z - y - \sqrt{2}\,t \) \] \, , \hspace{1cm} H_c := A \pi \cos \[ 2 \pi \( z - y - \sqrt{2}\,t \) \]  \, .
\eeq
The choice of parameters is again
\beq
T = 1000 \, , \hspace{1cm}  C = 0.25  \, , \hspace{1cm} A = 0.1 \, , \hspace{1cm} \si = 1 \, .
\eeq
For both the aligned and diagonal versions, we output the $L_{\infty}$ norm of the absolute errors $\De g_{tt}$, $\De g_{zz}$ and $\De g_{tz}$ and of the constraints $\Te$, $\ti{Z}^i$, $H$ and $M_i$. 

\begin{figure}
\includegraphics[width=\columnwidth]{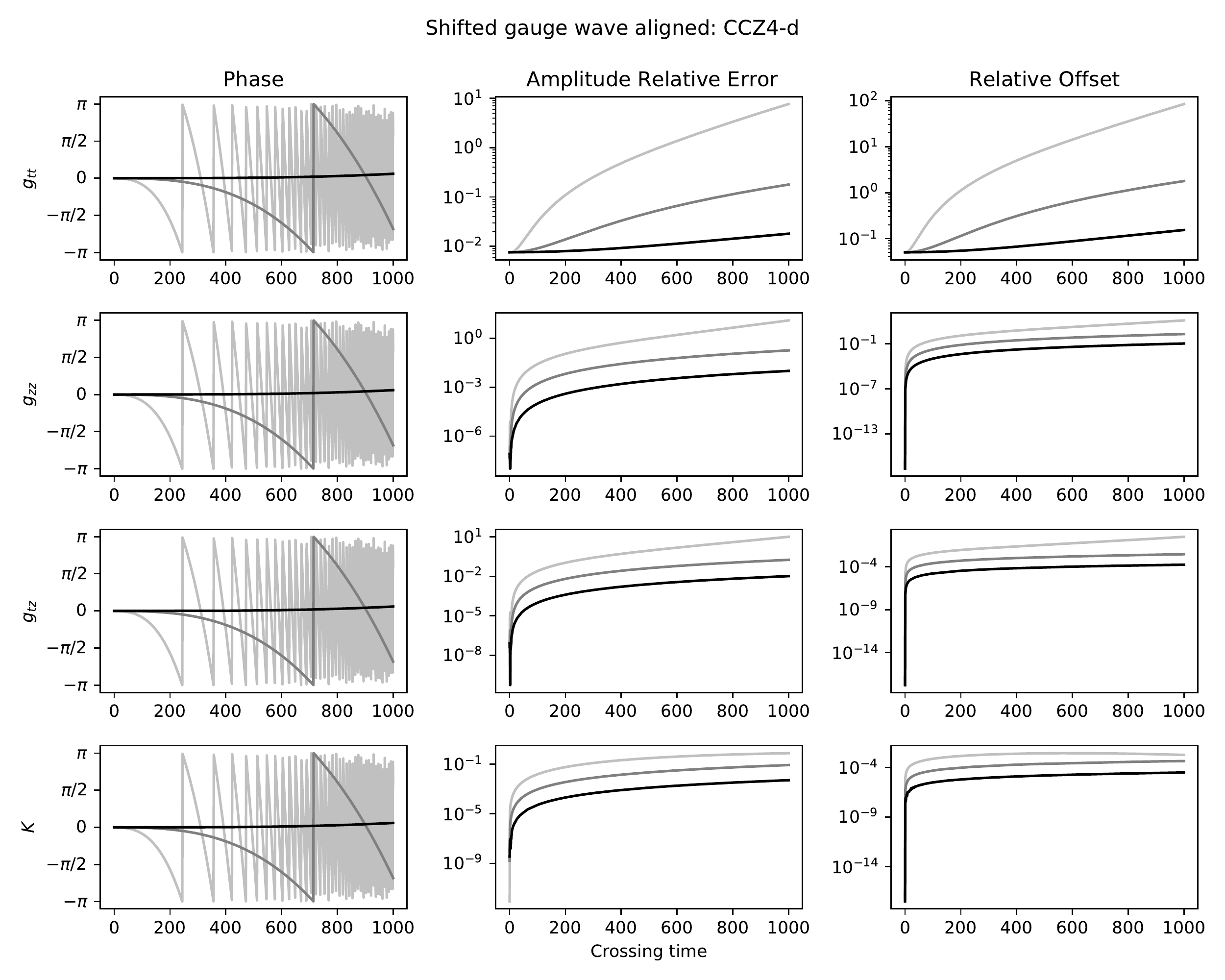} 
\caption{The relative phase ${\rm arg}\( F_1 \)$, relative error on the amplitude $\de F_1$ (``amplitude relative error") and the relative zero mode $\de F_0$ (``offset") of $g_{tt} + 1$, $g_{tz}$, $g_{zz} - 1$ and $K$ as a function of the number of crossing times in the aligned shifted gauge wave test for the CCZ4-d scheme. Each plot contains the three resolutions $\ro \in \{ 1,2,4 \}$ corresponding to the light-gray, dark-gray and black lines, respectively.}
\label{fig:ShiftGauge_CCZ4d_pao}
\end{figure}

\begin{figure}
\includegraphics[width=\columnwidth]{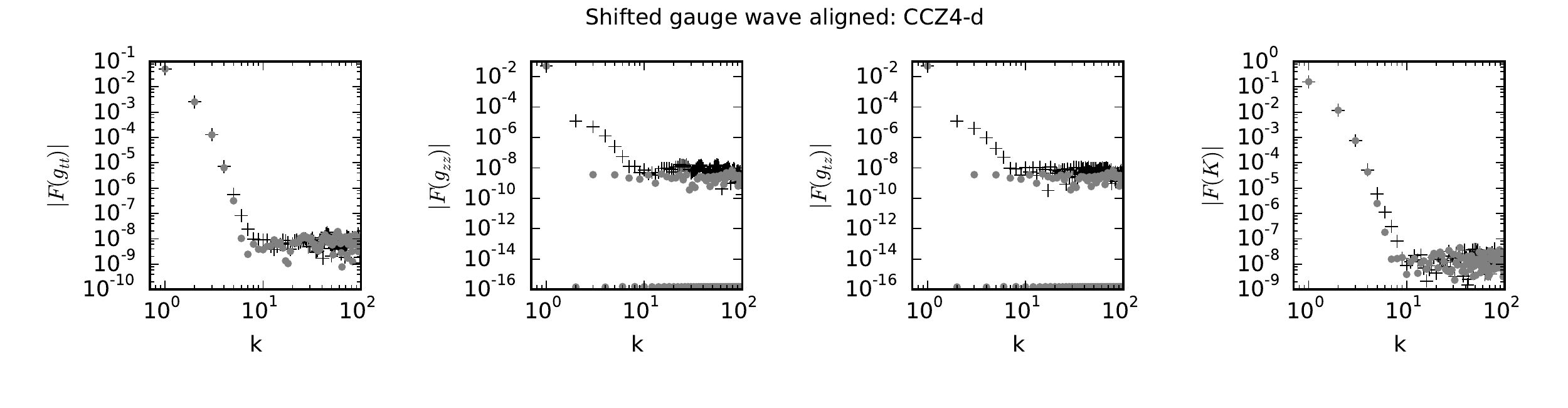} 
\caption{The modulus of the final spectra of $g_{tt} + 1$, $g_{tz}$, $g_{zz} - 1$ and $K$ in the aligned shifted gauge wave test for the CCZ4-d scheme at the highest resolution $\ro = 4$. The black crosses correspond to the numerical solution, while the gray circles correspond to the analytical solution.}
\label{fig:ShiftGauge_cc_Z4_pao}
\end{figure}

For the scheme that does not crash (CCZ4-d), we also monitor the spectrum of $g_{tt} + 1$, $g_{zz} - 1$, $g_{tz}$ and $K$ for the aligned version (see section \ref{sec:linewavespec} for a discussion of this measure). The quantities ${\rm arg}\( F_1\)$, $\de F_1$ and $\de F_0$ of the linearized wave test are defined here analogously for each aforementioned quantity and relatively to its typical analytical amplitude, and are displayed in fig. \ref{fig:ShiftGauge_CCZ4d_pao}. The modulus of the final spectrum for the highest resolution $\ro = 4$ is given in fig. \ref{fig:ShiftGauge_cc_Z4_pao}.

\subsection{Comments}

\begin{itemize}

\item The same comments as in the gauge wave test apply here as well.

\end{itemize}

\section{Gowdy wave test}

\subsection{Specifications}

For the ``expanding" solution, one uses the algebraic slicing condition 
\beq
\al = \sqrt{\ti{\ga}_{33}/\chi}  \, ,
\eeq
zero shift 
\beq
\be^i = 0 \, , 
\eeq
and 
\beq
\al = t^{-1/4} e^{\la/4} \, , \hspace{1cm} \chi = t^{-1/2} e^{- \la/6} \, , \hspace{1cm} K = -\frac{1}{4} \, t^{-3/4} e^{-\la/4} \( 3 + C \) \, , 
\eeq
\beq
\hat{\Ga}^{1,2} = 0 \, , \hspace{1cm}  \hat{\Ga}^3 = \frac{4 \pi^2}{3}\, t^2 e^{-\la/3} J_0(2\pi t) \, J_1(2\pi t) \, \sin(4\pi z)  \, , 
\eeq
\beq
\ti{\ga}_{ij} = \( \begin{array}{ccc} t^{1/2} e^{P - \la/6}  & 0  & 0 \\ 0  & t^{1/2} e^{-P - \la/6}  & 0 \\ 0 & 0 & t^{-1} e^{\la/3}  \end{array} \)  \, , 
\eeq
\beq
\ti{A}_{ij} = -\frac{1}{12} \( \begin{array}{ccc} t^{-1/4} e^{P-5 \la/12} \( 3 - C_+ \) & 0 & 0 \\ 0 & t^{-1/4} e^{-P-5\la/12} \( 3 - C_- \) & 0 \\ 0 & 0 & -2 t^{-7/4} e^{\la/12} \( 3 - C \) \end{array} \) \, ,
\eeq
where
\beq
P := J_0(2\pi t)\, \cos(2\pi z) \, ,
\eeq
\bea
\la & := & 2\pi^2 t^2 \[ J_0^2(2\pi t) + J_1^2(2\pi t) \] - 2\pi t J_0(2\pi t) J_1(2\pi t) \cos^2(2\pi z) \nn \\
 & & - \, 2\pi^2 \[ J^2_0(2\pi) + J^2_1(2\pi) \]  + \pi J_0(2\pi)\, J_1(2\pi) \, , \\
C_{\pm} & := & 4\pi t \[ J_1(2\pi t)  \cos(2\pi z) \( \pm 3 + \pi t J_1(2\pi t) \cos(2\pi z) \) + \pi t J_0^2(2\pi t) \sin^2(2\pi z) \] \, , \\
C & := & 4 \pi^2 t^2 \[ J_0^2(2\pi t) \sin^2(2\pi z) + J_1^2(2\pi t) \cos^2(2\pi z) \] \, .
\eea
and starts at $t_{\rm in.} = 1$. For the ``contracting" solution, one uses harmonic slicing, zero shift and
\beq
\al = - c k^{3/4} e^{-3ct/4 + \ti{\la}/4} \, , \hspace{1cm} \chi = k^{-1/2} e^{- \ti{\la}/6 + c t/2} \, , \hspace{1cm} K = -\frac{1}{4} \, k^{-3/4} e^{-\la/4+3ct/4} \( 3 + \ti{C} \) \, ,  
\eeq
\beq
\hat{\Ga}^{1,2} = 0 \, , \hspace{1cm}  \hat{\Ga}^3 = \frac{4 \pi^2}{3}\, k^2 e^{-2ct-\ti{\la}/3} J_0(2\pi k e^{-ct}) \, J_1(2\pi k e^{-ct}) \, \sin(4\pi z)  \, ,
\eeq
\beq
\ti{\ga}_{ij} = \( \begin{array}{ccc} k^{1/2} e^{\ti{P} - \ti{\la}/6 - ct/2} & 0  & 0 \\ 0  & k^{1/2} e^{-\ti{P} - \ti{\la}/6 - c t/2}  & 0 \\ 0 & 0 & k^{-1} e^{\ti{\la}/3 + ct}  \end{array} \)  \, , 
\eeq
\beq
\ti{A}_{ij} = -\frac{1}{12} \( \begin{array}{ccc} k^{-1/4} e^{\ti{P}-5 \ti{\la}/12 + c t/4} \( 3 - \ti{C}_+ \) & 0 & 0 \\ 0 & k^{-1/4} e^{-\ti{P}-5\ti{\la}/12 + ct/4} \( 3 - \ti{C}_- \) & 0 \\ 0 & 0 & -2k^{-7/4} e^{\ti{\la}/12 + 7ct/4} \( 3 - \ti{C} \) \end{array} \) \, ,
\eeq
where
\beq
\ti{P}(t,z) := P(k e^{-ct},z) \, , \hspace{0.5cm} \ti{\la}(t,z) := \la(k e^{-ct},z)  \, , \hspace{0.5cm} \ti{C}_{\pm}(t,z) := C_{\pm}(k e^{-ct},z)  \, , \hspace{0.5cm} \ti{C}(t,z) := C(k e^{-ct},z) \, ,
\eeq
\beq
c = 0.0021195119214617 \, , \hspace{1cm} k = 9.6707698127638 \, ,
\eeq
and starts at $t_{\rm in.} = -9.8753205829098$. The parameters are
\beq
T = 1000 \, , \hspace{1cm}  C = 0.25 \, , \hspace{1cm} \si = 0.05  \, .
\eeq
For the expanding case, we output the evolution of $L_{\infty}$ norm of the absolute errors $\De g_{tt}$, $\De g_{xx}$ and of the constraints $\Te$, $\ti{Z}^i$, $H$ and $M_i$. 

For the contracting case, we output the evolution of $L_{\infty}$ norm of the absolute errors $\De g_{tt}$, $\De g_{zz}$ and $\De g_{xx}$ and of the constraints $\Te$, $\ti{Z}^i$, $H$ and $M_i$. On top of this, we also display the profiles of $g_{zz}$, $g_{tt}$, $K$ and $|H|$ along the $z$ axis at $T = 1000$.

\subsection{Comments}

\begin{figure}
\includegraphics[width=\columnwidth]{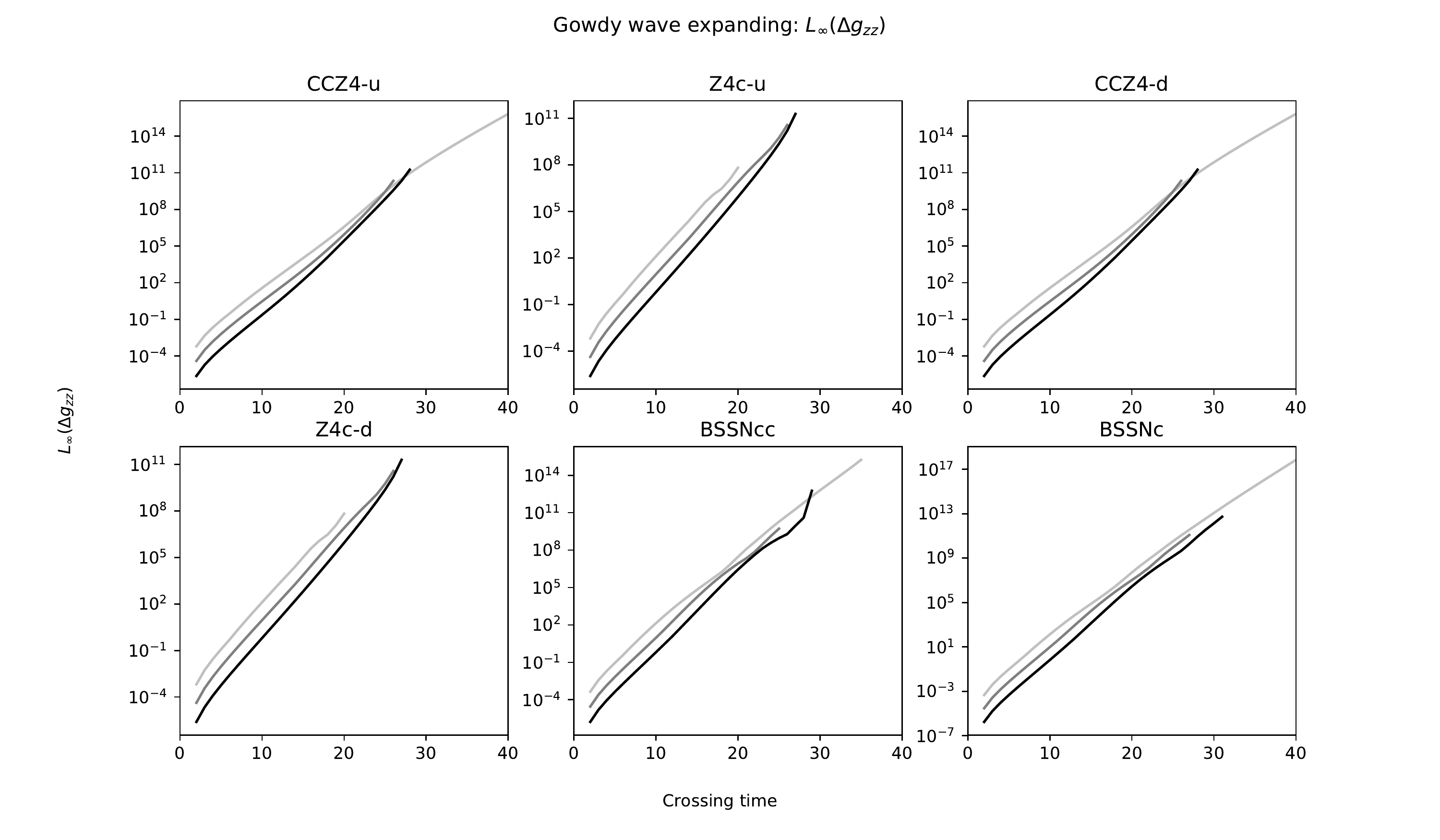} 
\caption{The $L_{\infty}$ norm of the absolute error $\De g_{zz}$ as a function of the number of crossing times in the expanding Gowdy wave test for all eight schemes. Each plot contains the three resolutions $\ro \in \{ 1,2,4 \}$ corresponding to the light-gray, dark-gray and black lines, respectively.}
\label{fig:Gowdy_Linf_error_gzz_expand}
\end{figure}

\begin{figure}
\includegraphics[width=\columnwidth]{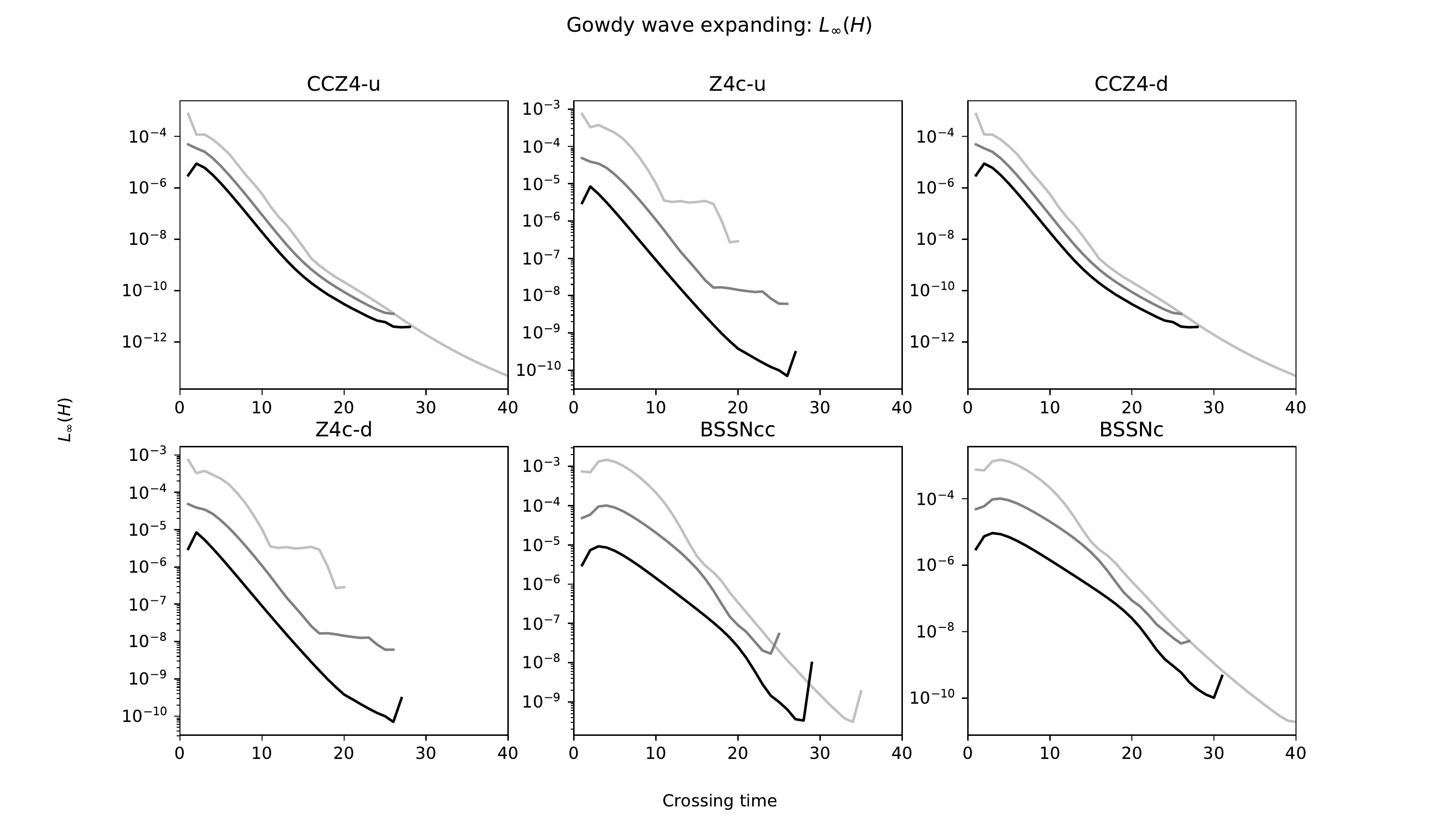} 
\caption{The $L_{\infty}$ norm of $H$ as a function of the number of crossing times in the expanding Gowdy wave test for all eight schemes. Each plot contains the three resolutions $\ro \in \{ 1,2,4 \}$ corresponding to the light-gray, dark-gray and black lines, respectively.}
\label{fig:Gowdy_Linf_H_expand}
\end{figure}

\begin{figure}
\includegraphics[width=\columnwidth]{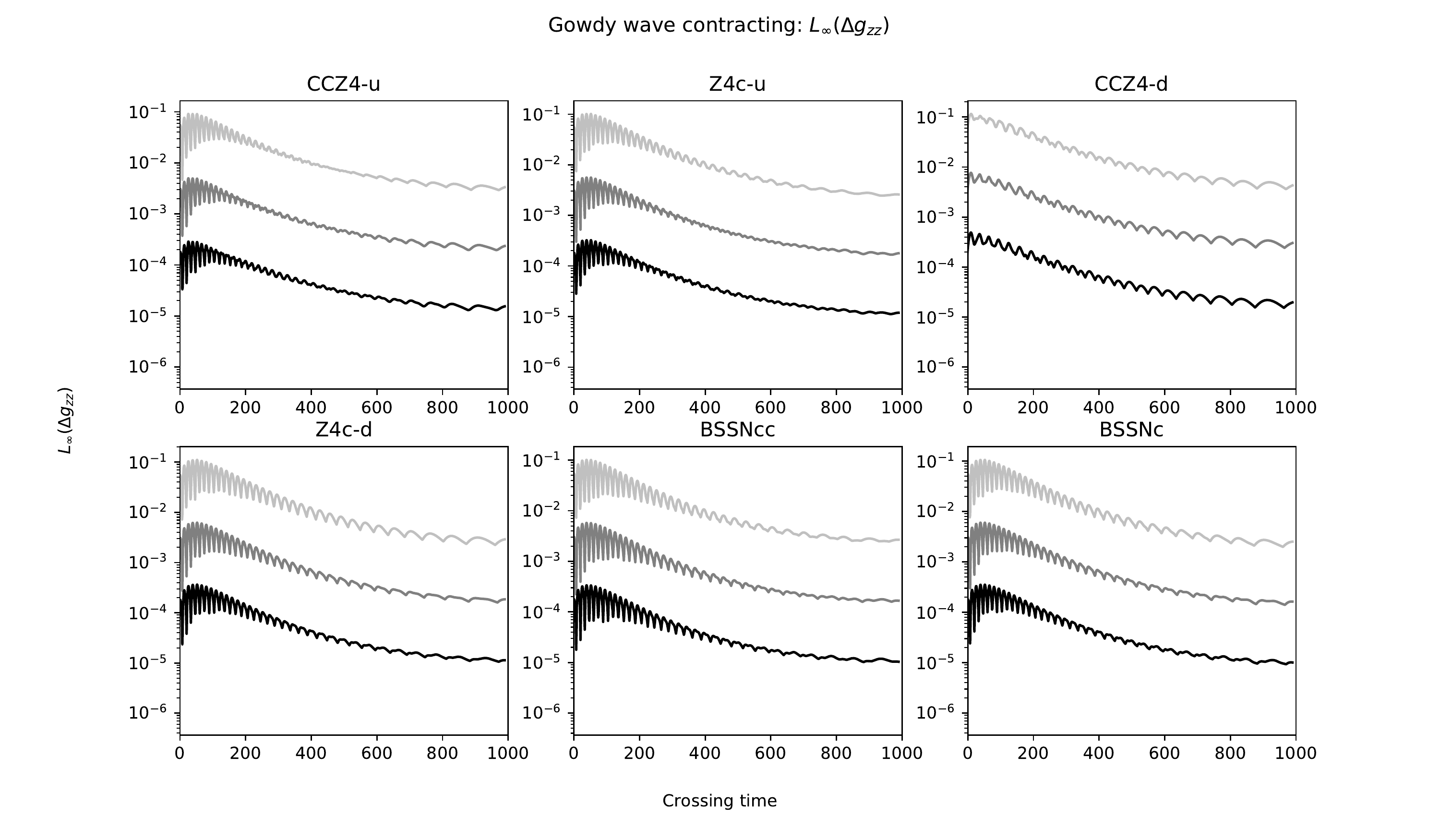} 
\caption{The $L_{\infty}$ norm of the absolute error $\De g_{zz}$ as a function of the number of crossing times in the contracting Gowdy wave test for all eight schemes. Each plot contains the three resolutions $\ro \in \{ 1,2,4 \}$ corresponding to the light-gray, dark-gray and black lines, respectively.}
\label{fig:Gowdy_Linf_error_gzz_contract}
\end{figure}

\begin{figure}
\includegraphics[width=\columnwidth]{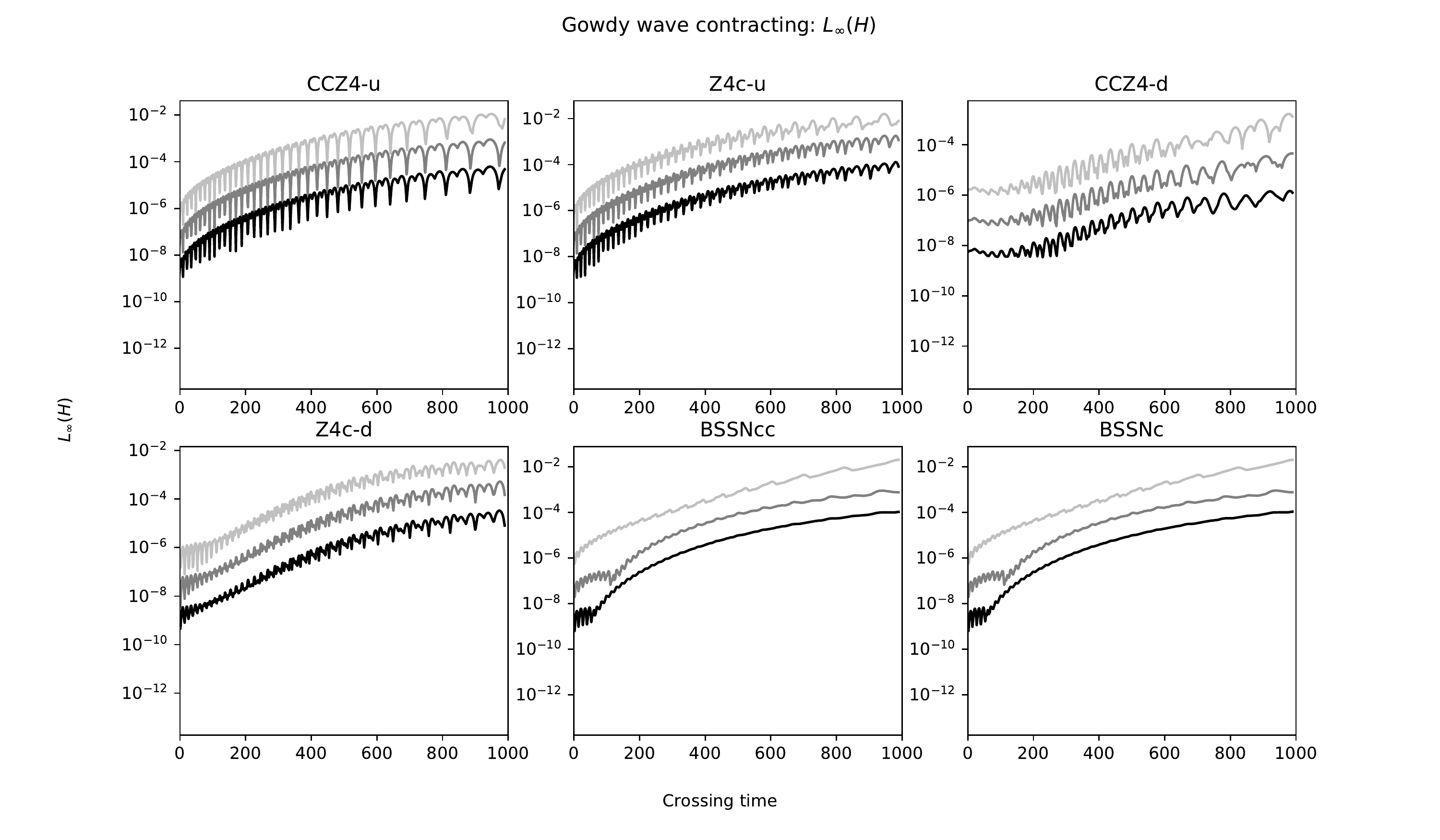} 
\caption{The $L_{\infty}$ norm of $H$ as a function of the number of crossing times in the contracting Gowdy wave test for all eight schemes. Each plot contains the three resolutions $\ro \in \{ 1,2,4 \}$ corresponding to the light-gray, dark-gray and black lines, respectively.}
\label{fig:Gowdy_Linf_H_contract}
\end{figure}

\begin{figure}
\center
\includegraphics[width=0.85\columnwidth]{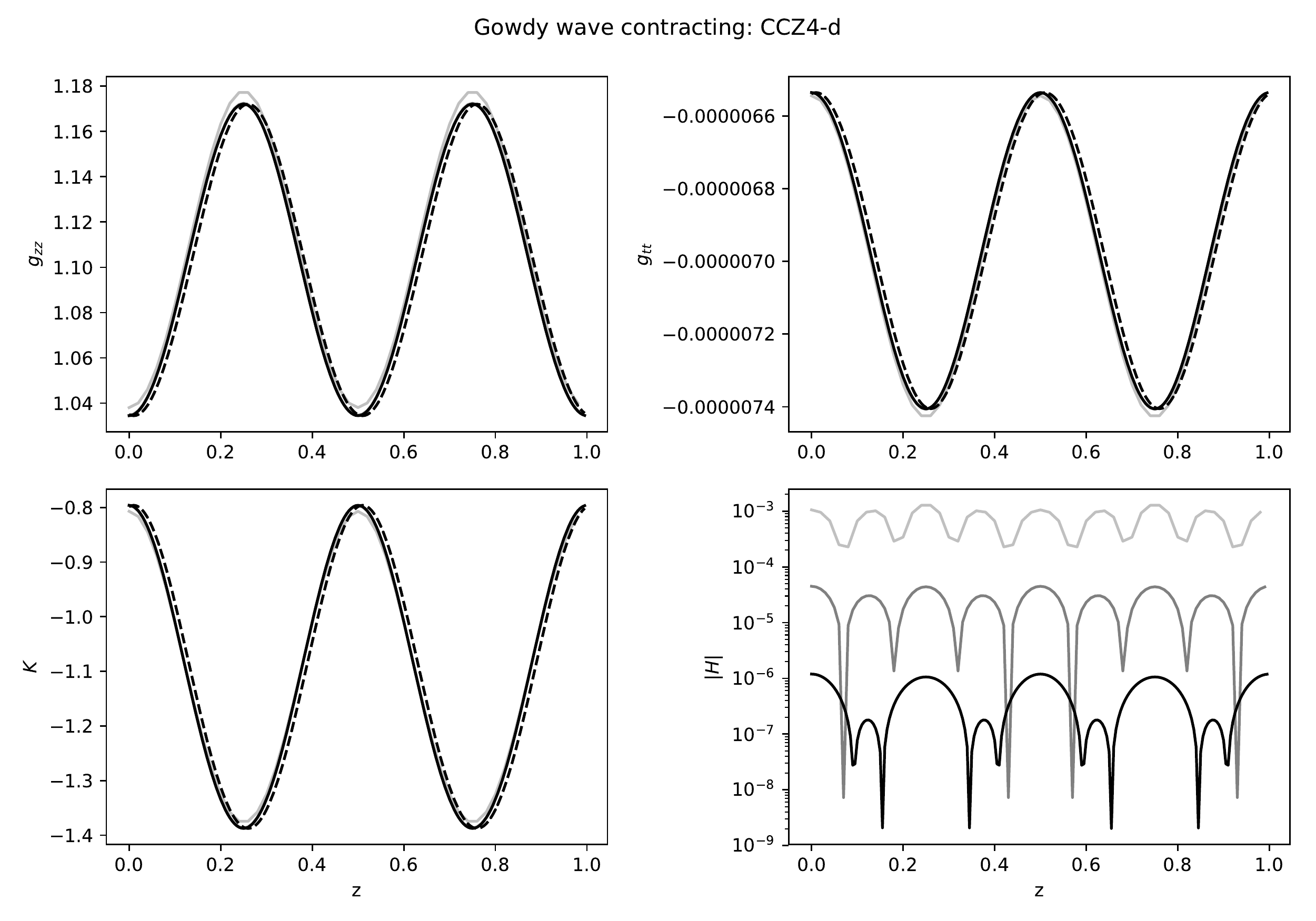} 
\caption{The profiles of $g_{zz}$, $g_{tt}$, $K$ and $|H|$ along the $z$ axis at $T = 1000$ for the CCZ4-d scheme. Each plot contains the three resolutions $\ro \in \{ 1,2,4 \}$ corresponding to the light-gray, dark-gray and black lines, respectively, while the dashed line corresponds to the analytical solution.}
\label{fig:Gowdy_CCZ4-d_wave}
\end{figure}

\begin{figure}
\center
\includegraphics[width=0.85\columnwidth]{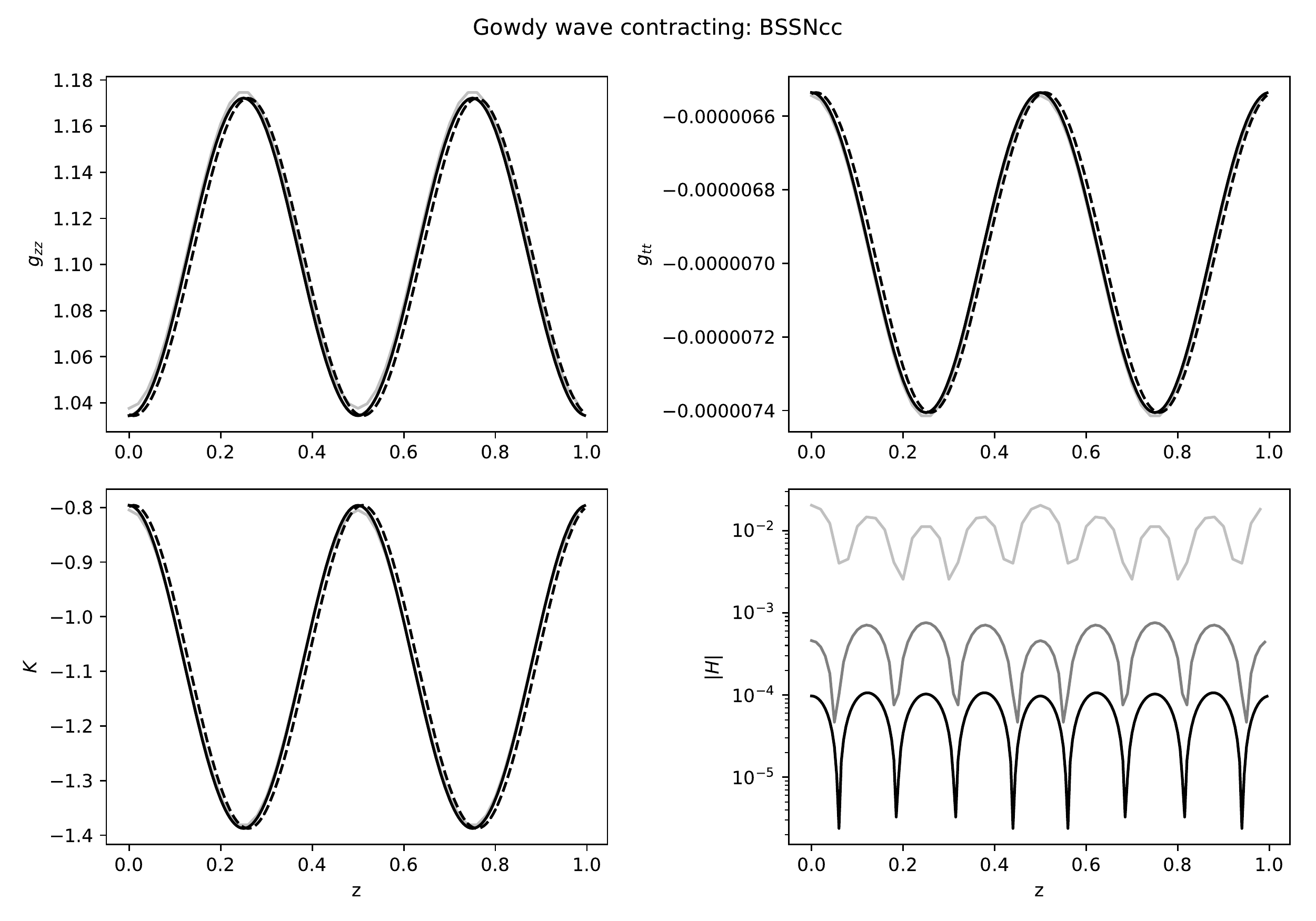} 
\caption{The profiles of $g_{zz}$, $g_{tt}$, $K$ and $|H|$ along the $z$ axis at $T = 1000$ for the BSSNcc scheme. Each plot contains the three resolutions $\ro \in \{ 1,2,4 \}$ corresponding to the light-gray, dark-gray and black lines, respectively, while the dashed line corresponds to the analytical solution.}
\label{fig:Gowdy_BSSNcc_wave}
\end{figure}

\begin{itemize}

\item
The conformal-covariant schemes CCZ4-d and Z4cc-d behave generically better than the others. In particular, the Hamiltonian constraint is one to two orders of magnitude lower.

\item
Expanding case (see fig. \ref{fig:Gowdy_Linf_error_gzz_expand} and \ref{fig:Gowdy_Linf_H_expand}):

\begin{itemize}
\item
We obtain the usual result \cite{A2A2, CFFKLT} that the run crashes at around $T \approx 30$ crossing times.
\item 
All schemes lead qualitatively to the same results. In particular, damping has no significant impact. Nevertheless, the Z4 schemes perform better than BSSN.
\item
The $H$ and $M_i$ constraints are bounded in all schemes.
\end{itemize}

\item
Contracting case (see figs. \ref{fig:Gowdy_Linf_error_gzz_contract} to \ref{fig:Gowdy_BSSNcc_wave}):

\begin{itemize}

\item 
The errors are comparable between the schemes.

\item 
Constraint violations increase by a few orders of magnitude during the simulation and damping $\ka > 0$ reduces that violation in the course of the simulation.

\item 
At $T = 1000$ the wave profiles are visibly close to the analytical solution and the Hamiltonian constraint profile seems to converge uniformly.

\end{itemize}

\end{itemize}

\acknowledgments

This work has been supported by several grants of the Swiss National Science Foundation for DD and YD, by a Consolidator Grant of the European Research Council (ERC-2015-CoG grant 680886) for YD and EM and by a grant from the Swiss National Supercomputing Centre (CSCS) under project ID s751. The simulations have been performed on the Baobab cluster of the University of Geneva, on Piz Daint of the CSCS and on the Cambridge COSMOS SMP system (part of the STFC DiRAC HPC Facility supported by BIS NeI capital grant ST/J005673/1 and STFC grants ST/H008586/1, ST/K00333X/1).

\clearpage



%

\clearpage



\clearpage


\end{document}

%% file: mydefs.tex
\newcommand{\beq}{\begin{equation}}
\newcommand{\eeq}{\end{equation}}
\newcommand{\bea}{\begin{eqnarray}}
\newcommand{\eea}{\end{eqnarray}}
\newcommand{\ben}{\begin{enumerate}}
\newcommand{\een}{\end{enumerate}}

\newcommand{\pa}{\partial}

\newcommand{\na}{\nabla}
\newcommand{\ed}{{\rm d}}
\newcommand{\ced}{{\rm D}}

\newcommand{\ti}{\tilde}

\renewcommand\({\left(}
\renewcommand\){\right)}
\renewcommand\[{\left[}
\renewcommand\]{\right]}

\newcommand{\nn}{\nonumber}

\newcommand{\al}{\alpha}
\newcommand{\be}{\beta}
\newcommand{\ga}{\gamma}
\newcommand{\Ga}{\Gamma}
\newcommand{\de}{\delta}
\newcommand{\De}{\Delta}
\newcommand{\ep}{\epsilon}

\newcommand{\Te}{\Theta}

\newcommand{\ka}{\kappa}
\newcommand{\la}{\lambda}

\newcommand{\ro}{\rho}
\newcommand{\si}{\sigma}

\newcommand{\Lie}{{\cal L}}

%% file: Main.bbl
\begin{thebibliography}{20}   

\bibitem{HF}
  H.~Friedrich, ``On the hyperbolicity of Einstein's and other gauge field equations'',
  Communications\ in\ Mathematical\ Physics\  {\bf 100} (1985) 4.

\bibitem{Garfinkle} 
  D.~Garfinkle, ``Harmonic coordinate method for simulating generic singularities'',
  Phys.\ Rev.\ D {\bf 65} (2002) 044029, \href{http://xxx.lanl.gov/abs/gr-qc/0110013}{{\tt gr-qc/0110013}}.

\bibitem{SW} 
  B.~Szilagyi and J.~Winicour,  ``Well posed initial boundary evolution in general relativity'',
  Phys.\ Rev.\ D {\bf 68} (2003) 041501, \href{http://xxx.lanl.gov/abs/gr-qc/0205044}{{\tt gr-qc/0205044}}.

\bibitem{Pretorius}
  F.~Pretorius, ``Numerical relativity using a generalized harmonic decomposition'', 
  Class.\ Quant.\ Grav.\  {\bf 22} (2005) 425, \href{http://xxx.lanl.gov/abs/gr-qc/0407110}{{\tt gr-qc/0407110}}.
  
 \bibitem{LSKOR}
 L.~Lindblom, M.~A.~Scheel, L.~E.~Kidder, R.~Owen, O.~Rinne, `` A New generalized harmonic evolution system'', 
  Class.\ Quant.\ Grav.\  {\bf 23} (2006) 447, \href{http://xxx.lanl.gov/abs/gr-qc/0512093}{{\tt gr-qc/0512093}}.

\bibitem{NOK}
  T.~Nakamura, K.~Oohara and Y.~Kojima,  ``General Relativistic Collapse to Black Holes and Gravitational Waves from Black Holes'', 
  Prog.\ Theor.\ Phys.\ Suppl.\  {\bf 90} (1987) 1.

\bibitem{SN} 
  M.~Shibata and T.~Nakamura,  ``Evolution of three-dimensional gravitational waves: Harmonic slicing case'', 
  Phys.\ Rev.\ D {\bf 52}, 5428 (1995). 

\bibitem{BS} 
  T.~W.~Baumgarte and S.~L.~Shapiro,   ``On the numerical integration of Einstein's field equations'', 
  Phys.\ Rev.\ D {\bf 59}, 024007 (1998), \href{http://xxx.lanl.gov/abs/gr-qc/9810065}{{\tt gr-qc/9810065}}.

\bibitem{BLPZ}
  C.~Bona, T.~Ledvinka, C.~Palenzuela and M.~Zacek,   ``General covariant evolution formalism for numerical relativity'',
  Phys.\ Rev.\ D {\bf 67} (2003) 104005, \href{http://xxx.lanl.gov/abs/gr-qc/0302083}{{\tt gr-qc/0302083}}. 

\bibitem{GMGCH}
  C.~Gundlach, J.~M.~Martin-Garcia, G.~Calabrese and I.~Hinder,   ``Constraint damping in the Z4 formulation and harmonic gauge'',
  Class.\ Quant.\ Grav.\  {\bf 22} (2005) 3767, \href{http://xxx.lanl.gov/abs/gr-qc/0504114}{{\tt gr-qc/0504114}}. 


\bibitem{BH}
  S.~Bernuzzi and D.~Hilditch,   ``Constraint violation in free evolution schemes: Comparing BSSNOK with a conformal decomposition of Z4'', 
  Phys.\ Rev.\ D {\bf 81} (2010) 084003, \href{http://xxx.lanl.gov/abs/0912.2920}{{\tt 0912.2920}}. 

\bibitem{WBH}
  A.~Weyhausen, S.~Bernuzzi and D.~Hilditch,  ``Constraint damping for the Z4c formulation of general relativity'',
  Phys.\ Rev.\ D {\bf 85} (2012) 024038, \href{http://xxx.lanl.gov/abs/1107.5539}{{\tt 1107.5539}}.   


\bibitem{ABCBRP}
  D.~Alic, C.~Bona-Casas, C.~Bona, L.~Rezzolla and C.~Palenzuela,   ``Conformal and covariant formulation of the Z4 system with constraint-violation damping'',
  Phys.\ Rev.\ D {\bf 85} (2012) 064040, \href{http://xxx.lanl.gov/abs/1106.2254}{{\tt 1106.2254}}. 


\bibitem{A2A1} 
  M.~Alcubierre {\it et al.},   ``Toward standard testbeds for numerical relativity'', 
  Class.\ Quant.\ Grav.\  {\bf 21}, 589 (2004), \href{http://xxx.lanl.gov/abs/gr-qc/0305023}{{\tt gr-qc/0305023}}.
  

\bibitem{A2A2} 
  M.~C.~Babiuc {\it et al.},   ``Implementation of standard testbeds for numerical relativity'',
  Class.\ Quant.\ Grav.\  {\bf 25}, 125012 (2008),  \href{http://xxx.lanl.gov/abs/0709.3559}{{\tt 0709.3559}}.  


\bibitem{CH}
  Z.~Cao and D.~Hilditch,   ``Numerical stability of the Z4c formulation of general relativity'',
  Phys.\ Rev.\ D {\bf 85} (2012) 124032, \href{http://xxx.lanl.gov/abs/1111.2177}{{\tt 1111.2177}}.


\bibitem{DGKRZ}
  M.~Dumbser, F.~Guercilena, S.~Koeppel, L.~Rezzolla, O.~Zanotti, ``Conformal and covariant Z4 formulation of the Einstein equations: strongly hyperbolic first-order reduction and solution with discontinuous Galerkin schemes'',
  Phys.\ Rev.\ D\  {\bf 97} (2018) no.8, 084053, \href{http://xxx.lanl.gov/abs/1707.09910}{{\tt 1707.09910}}.

\bibitem{MGS}
  J.~B.~Mertens, J.~T.~Giblin and G.~D.~Starkman, ``Integration of inhomogeneous cosmological spacetimes in the BSSN formalism'',
  Phys.\ Rev.\ D {\bf 93} (2016) no.12,  124059, \href{http://xxx.lanl.gov/abs/1511.01106}{{\tt 1511.01106}}.


\bibitem{CFFKLT}
  K.~Clough, P.~Figueras, H.~Finkel, M.~Kunesch, E.~A.~Lim and S.~Tunyasuvunakool, ``GRChombo : Numerical Relativity with Adaptive Mesh Refinement'',
  Class.\ Quant.\ Grav.\  {\bf 32} (2015) no.24,  245011, \href{http://xxx.lanl.gov/abs/1503.03436}{{\tt 1503.03436}}.

\bibitem{BMSS}
  C.~Bona, J.~Masso, E.~Seidel and J.~Stela, ``A New formalism for numerical relativity'',
  Phys.\ Rev.\ Lett.\  {\bf 75} (1995) 600, \href{http://xxx.lanl.gov/abs/gr-qc/9412071}{{\tt gr-qc/9412071}}.


\bibitem{BCCKvM}
  J.~G.~Baker, J.~Centrella, D.~I.~Choi, M.~Koppitz and J.~van Meter, ``Gravitational wave extraction from an inspiraling configuration of merging black holes'',
  Phys.\ Rev.\ Lett.\  {\bf 96} (2006) 111102, \href{http://xxx.lanl.gov/abs/gr-qc/0511103}{{\tt gr-qc/0511103}}.


\bibitem{CLMZ}
  M.~Campanelli, C.~O.~Lousto, P.~Marronetti and Y.~Zlochower, ``Accurate evolutions of orbiting black-hole binaries without excision'',
  Phys.\ Rev.\ Lett.\  {\bf 96} (2006) 111101, \href{http://xxx.lanl.gov/abs/gr-qc/0511048}{{\tt gr-qc/0511048}}.


\bibitem{BB}
  S.~Brandt and B.~Bruegmann, ``A Simple construction of initial data for multiple black holes'',
  Phys.\ Rev.\ Lett.\  {\bf 78} (1997) 3606, \href{http://xxx.lanl.gov/abs/gr-qc/9703066}{{\tt gr-qc/9703066}}.

\bibitem{KO}
H.~O.~Kreiss and J.~Oliger, ``Methods for the approximate solution of time dependent problems", GARP publication series No. 10, Geneva, 1973.

\bibitem{Alcubierre}
M.~Alcubierre, ``Introduction to 3+1 numerical relativity", 
International Series of Monographs on Physics 140, (Oxford University Press, 2008).

  




 
\end{thebibliography}
